\documentclass{article}
\usepackage{graphicx} 
\usepackage{authblk}
\usepackage[english]{babel}
\usepackage{amsmath}
\usepackage{amssymb}
\usepackage{caption}
\usepackage{subcaption}
\usepackage{comment}
\usepackage{xcolor}
\usepackage{csquotes}
\usepackage{etoc}

\definecolor{linkblue}{RGB}{0,70,140}
\usepackage[colorlinks=true, citecolor=linkblue, linkcolor=linkblue, urlcolor=linkblue]{hyperref}

\usepackage[backend=biber, style=authoryear, backref=true, sorting=nyt, uniquename=false]{biblatex}
\definecolor{haroldblue}{HTML}{1F78B4}

\providecommand{\keywords}[1]
{
  \small	
  \textbf{\textit{Keywords:}} #1
} 

\usepackage{color}
\usepackage{tikz}
\usetikzlibrary{shapes,decorations,arrows,calc,arrows.meta,fit,positioning}
\tikzset{
    -Latex,auto,node distance =1 cm and 1 cm,semithick,
    state/.style ={ellipse, draw, minimum width = 0.7 cm},
    point/.style = {circle, draw, inner sep=0.04cm,fill,node contents={}},
    bidirected/.style={Latex-Latex,dashed},
    el/.style = {inner sep=2pt, align=left, sloped}
}

\newcommand{\indep}{{\rotatebox[origin=c]{90}{$\models$}}}

\title{Partitioning Time in Target Trial Emulation}
\author[1,*]{Harold~Tankpinou Zoumenou}
\author[1]{Simon~Ferreira}
\author[1]{Charles~Assaad}
\author[2]{David~Hajage}
\author[3]{Alexandra~Beurton}
\author[3]{Fabrice~Carrat}
\author[3]{Nathanaël~Lapidus}
\author[1]{Daria~Bystrova}
\author[3,4]{Benjamin~Glemain}
\affil[1]{Sorbonne Université, Inserm, Sorbonne Public Health Institute, F75012 Paris, France}
\affil[2]{Sorbonne Université, Inserm, Sorbonne Public Health Institute, Equipe PEPITES, Assistance Publique - Hôpitaux de Paris, Hôpital Pitié Salpêtrière, Département de Santé Publique, Centre de Pharmacoépidémiologie (Cephepi), Paris, France.}
\affil[3]{Sorbonne Université, Inserm, Sorbonne Public Health Institute, APHP, Hôpital Saint-Antoine, Département de Santé Publique, F75012 Paris, France}
\affil[4]{Sorbonne Université, Inserm, Inria, CNRS, Paris Brain Institute, Paris, France}
\affil[*]{Corresponding author. E-mail: harold.tankpinou@iplesp.upmc.fr}
\date{\today}

\begin{document}

\maketitle

\begin{abstract}

In target trial emulation, the treatment strategies that patients follow are inferred from the treatments they actually receive in routine care. However, in most settings, the outcome may preclude the observation of planned treatment, giving rise to immortal time bias through misclassification of treatment strategy, while markers of treatment response may influence subsequent treatment decisions, giving rise to time-varying confounding. A key step toward unbiased treatment effect estimation is to partition follow-up into sufficiently short time intervals to unfold the feedback relationships involving treatment and represent the resulting causal relations with a directed acyclic graph. In this study, we present the possible within-interval causal orderings induced by this partitioning, discuss their causal implications, and assess their plausibility across clinical settings. For each causal ordering, we derive the corresponding g-formula. Using ancestral multi-world networks and simulations, we show that the standard cloning-censoring-weighting estimator is invalid when treatment affects the outcome within a time interval, and we propose a modified version of the method that restores its validity in this setting. Finally, we analyze the consequences of choosing time intervals that are either excessively wide or excessively narrow, thereby formally establishing the need for time partitioning and providing practical guidance for selecting an appropriate partition based on the clinical setting.

\end{abstract}


\section{Introduction}

In randomized controlled trials, randomization ensures comparability between treatment groups with respect to both measured and unmeasured confounders. Furthermore, treatment allocation is known for all participants, including those who die before receiving their assigned treatment.

Target trial emulation, based on the explicit specification of a hypothetical randomized trial to eliminate any ambiguity regarding the effect under investigation (study population, treatment strategies, outcome), consists of estimating the effect of a treatment using observational data  \parencite{hernanUsingBigData2016}. In such observational settings, a patient intended to receive a treatment but dies before its initiation may be classified in the control group, whereas the same patient would have been classified in the treatment group had they survived long enough to initiate treatment. This creates a circular relationship in which early death results in treatment misclassification, thereby giving rise to immortal time bias \parencite{hernanStructuralDescriptionBiases2025}.
More generally, immortal time bias may arise for any non-fatal outcome when treatment strategy is defined based only on treatment received prior to outcome occurrence. Another form of causal feedback may arise between treatment and markers of treatment response, such as biological monitoring measures, as these markers influence subsequent treatment decisions, leading to time-dependent confounding. For example, a patient untreated at baseline may initiate treatment during follow-up because of worsening biological markers.

Immortal time bias and time-dependent confounding can be avoided by comparing treatment strategies that are fully determined at the start of follow-up (also referred to as ``time zero"), such as in studies comparing surgical procedures performed immediately after inclusion \parencite{hernanStructuralDescriptionBiases2025, guillot-tantayLongtermSafetyMidurethral2025}. In such settings, the observed treatment strategy cannot be influenced by subsequent death or markers of treatment response. In all other cases, breaking the apparent causal cycles between treatment and death or between treatment and time-varying confounders requires partitioning follow-up time into smaller intervals, a necessary step toward unbiased estimation of treatment effects.

If partitioning results in sufficiently small time intervals, a directed \textit{acyclic} graph (DAG) can be used to represent the causal relationships between treatment, outcome, and confounders over time \parencite{pearlCausalDiagramsEmpirical1995a, pearlCausalityModelsReasoning2009, greenlandCausalDiagramsEpidemiologic1999}. With variables in a given time interval preceding those in subsequent intervals, designing a DAG requires specifying the within-interval causal ordering. The methodological literature describes several within-interval causal orderings that differ, in particular, in the assumed temporal ordering of treatment and outcome \parencite{robinsMarginalStructuralModels2000, hernanMarginalStructuralModels2000, mcgrathGfoRmulaPackageEstimating2020, wanisGracePeriodsComparative2024}.\footnote{In \cite{robinsMarginalStructuralModels2000} and in \cite{hernanMarginalStructuralModels2000}, covariates precede treatment and the pooled logistic regressions used to estimate hazards include treatment from previous time intervals only, suggesting that, within each interval, the outcome precedes treatment. Similarly, the first supplementary file of \cite{mcgrathGfoRmulaPackageEstimating2020} states that covariates precede treatment and that death is included among the covariates, which also implies that the outcome precedes treatment. By contrast, \cite{wanisGracePeriodsComparative2024} explicitly specifies a within-interval causal ordering in which covariates precede treatment and treatment precedes outcome.} To our knowledge, the plausibility of these causal structures across different clinical contexts has not been examined, nor has their interchangeability with respect to estimation procedures been evaluated. Clarifying these issues is important for target trial emulation. Indeed, reporting guidelines require that any causal assumptions needed to estimate causal effects be explicitly stated \parencite{cashinTransparentReportingObservational2025}.

This study aimed to clarify the challenges associated with time partitioning in target trial emulation and to provide practical guidance for addressing them. Section~\ref{sec:framework} presents the setting, notations, and causal framework used throughout the article. In Section~\ref{sec:causal_ordering}, we present two within-interval causal orderings that serve as references for the analyses developed in this article. We present their corresponding directed acyclic graphs (DAGs) and discuss the plausibility of their implications across several clinical settings. In Section~\ref{sec:estimation}, we evaluate the validity of nonparametric g-computation and cloning-censoring-weighting (CCW) under the two reference causal orderings \parencite{robinsNewApproachCausal1986, cainWhenStartTreatment2010}. Using ancestral multi-world networks (AMWNs) and simulations \parencite{correaCounterfactualGraphicalModels2025}, we show that CCW may be biased when treatment precedes outcome within time intervals and propose a modification of the method to address this bias. We also show that alternative within-interval causal orderings are equivalent, for estimation purposes, to the reference ordering that shares the same treatment-outcome ordering. Section~\ref{sec:practical_considerations} concludes the article by examining the consequences of choosing time intervals that are either excessively wide or excessively narrow, thereby formally establishing the need for time partitioning and providing practical guidance for selecting an appropriate partition based on the clinical setting.

\section{Framework and notation} \label{sec:framework}

This article examines time partitioning in target trial emulation \parencite{hernanUsingBigData2016}. Using observational data, we aim to estimate the causal effects of treatments on survival without meeting the absorbing outcome by the end of the study. To focus on time partitioning, this article considers static deterministic treatment strategies, in which treatment decisions do not depend on covariate history and a given strategy always yields the same treatment decisions \parencite{hernanCausalInferenceWhat2020}. For example, we consider treatment strategies in which patients either receive treatment throughout follow-up (for sustained treatments) or during the first study interval only (for one-time treatments, e.g., surgical procedures). These treatment strategies are compared with a control strategy in which patients never receive treatment.

$X$ denotes treatment, $Z$ denotes time-varying confounders, and $Y$ denotes the outcome. We consider the study duration to be partitioned into time intervals indexed from $1$ to $T$ (for a given patient, the first period begins at time zero, when the inclusion criteria are met). When a variable is represented over time, a subscript will indicate the period to which it refers. For example, $X_2$ denotes the treatment received during the second period of the study. An overbar denotes a variable and its history. For example, to denote all variables $Y_t$ for $t=1, \dots, T$, we write $\bar{Y}_{T}$. Uppercase letters represent random variables and lowercase letters represent particular realizations of those variables. For example, the expression $P(y_4 | \bar{X}_{3}=1)$ denotes the probability of observing a particular value $y_4$ for the variable $Y_4$ in patients who receive treatment in the first three time intervals.

Our study relies on the structural causal model (SCM) framework, which popularized the use of DAGs~\parencite{pearlCausalDiagramsEmpirical1995a, pearlCausalityModelsReasoning2009}. In a DAG, direct causal relationships between variables are encoded through directed edges from causes to effects. The d-separation criterion provides a graphical rule for identifying conditional independence relations implied by the DAG \parencite{geigerIdentifyingIndependenceBayesian1990}. Another central feature of the SCM framework is the do-operator, which represents an intervention on a variable. For example, the expression $P(Y_T=1 | do(\bar{X}_{T}=1))$ denotes the probability of surviving through the whole study under an intervention in which the entire population receives treatment throughout follow-up. This expression is equivalent to the counterfactual quantity $P(Y_T^{\bar{X}_{T}=1}=1)$. In contrast, the expression $P(Y_T=1 | \bar{X}_{T}=1)$ denotes the probability of survival when the analysis is restricted to individuals in the target population who receive treatment throughout follow-up in routine practice.

\section{Reference causal ordering} \label{sec:causal_ordering}

\subsection{Directed acyclic graphs}

In this section, we introduce two reference within-interval causal orderings, along with their corresponding DAGs and clinical implications. In Section~\ref{sec:estimation}, we show that alternative orderings can be reduced, for estimation purposes, to one of these two reference orderings.

In both reference orderings, written ($Z_t$, $Y_t$, $X_t$) and $(Z_t, X_t, Y_t)$, time varying confounders $Z_t$ precede both treatment $X_t$ and outcome $Y_t$ for any time interval $t$. $Z_1$ additionally includes baseline confounders. The two orderings differ only in the relative position of treatment and outcome, with $X_t$ occurring either before or after $Y_t$. DAGs illustrating this difference for a study with two periods are shown in Figure~\ref{fig:referrence_orderings}.

\begin{figure}
     \centering
     \begin{subfigure}[b]{.9\textwidth}
         \centering
\begin{tikzpicture}
\node[state, text={rgb:red,39;green,82;blue,184}, draw={rgb:red,39;green,82;blue,184}, fill=white] at (-9.30, 10.50) (X1) {$X_{1}$};
\node[state, text={rgb:red,39;green,82;blue,184}, draw={rgb:red,39;green,82;blue,184}, fill=white] at (-6.00, 10.50) (X2) {$X_{2}$};
\node[state, text=black, draw=black, fill=white] at (-9.30, 7.50) (Y1) {$Y_{1}$};
\node[state, text={rgb:red,230;green,26;blue,25}, draw={rgb:red,230;green,26;blue,25}, fill=white] at (-6.00, 7.50) (Y2) {$Y_{2}$};
\node[state, text=black, draw=black, fill=white] at (-11.10, 9.00) (Z1) {$Z_{1}$};
\node[state, text=black, draw=black, fill=white] at (-7.80, 9.00) (Z2) {$Z_{2}$};
\path (Y1) edge [bend left=0] (Y2);
\path (X1) edge [bend left=0] (X2);
\path[line width=2.5] (Y1) edge [bend left=0] (X1);
\path (Y1) edge [bend left=-24] (X2);
\path[line width=2.5] (Y2) edge [bend left=0] (X2);
\path (X1) edge [bend left=22] (Y2);
\path (Z1) edge [bend left=0] (X1);
\path (Z1) edge [bend left=0] (Y1);
\path (Z1) edge [bend left=-6] (X2);
\path (Z1) edge [bend left=0] (Y2);
\path (Z2) edge [bend left=0] (X2);
\path (Z2) edge [bend left=0] (Y2);
\path (X1) edge [bend left=0] (Z2);
\path (Z1) edge [bend left=0] (Z2);
\path (Y1) edge [bend left=0] (Z2);
\node[draw=black,dotted,fit=(Z1) (X1) (Y1), inner sep=0.2cm] (box1) {};
\node[above=2pt of box1, align=center] {1st time interval};
\node[draw=black,dotted,fit=(Z2) (X2) (Y2), inner sep=0.2cm] (box2) {};
\node[above=2pt of box2, align=center] {2nd time interval};
\end{tikzpicture}
         \caption{Reference causal ordering ($Z_t$, $Y_t$, $X_t$), where the outcome precedes treatment within time intervals}
         \vspace{.5cm}
         \label{fig:yt_xt}
     \end{subfigure}
     \begin{subfigure}[b]{.9\textwidth}
         \centering
\begin{tikzpicture}
\node[state, text={rgb:red,39;green,82;blue,184}, draw={rgb:red,39;green,82;blue,184}, fill=white] at (-9.30, 10.50) (X1) {$X_{1}$};
\node[state, text={rgb:red,39;green,82;blue,184}, draw={rgb:red,39;green,82;blue,184}, fill=white] at (-6.00, 10.50) (X2) {$X_{2}$};
\node[state, text=black, draw=black, fill=white] at (-9.30, 7.50) (Y1) {$Y_{1}$};
\node[state, text={rgb:red,230;green,26;blue,25}, draw={rgb:red,230;green,26;blue,25}, fill=white] at (-6.00, 7.50) (Y2) {$Y_{2}$};
\node[state, text=black, draw=black, fill=white] at (-11.10, 9.00) (Z1) {$Z_{1}$};
\node[state, text=black, draw=black, fill=white] at (-7.80, 9.00) (Z2) {$Z_{2}$};
\path (Y1) edge [bend left=0] (Y2);
\path (X1) edge [bend left=0] (X2);
\path[line width=2.5] (X1) edge [bend left=0] (Y1);
\path (Y1) edge [bend left=-24] (X2);
\path[line width=2.5] (X2) edge [bend left=0] (Y2);
\path (X1) edge [bend left=22] (Y2);
\path (Z1) edge [bend left=0] (X1);
\path (Z1) edge [bend left=0] (Y1);
\path (Z1) edge [bend left=-6] (X2);
\path (Z1) edge [bend left=0] (Y2);
\path (Z2) edge [bend left=0] (X2);
\path (Z2) edge [bend left=0] (Y2);
\path (X1) edge [bend left=0] (Z2);
\path (Z1) edge [bend left=0] (Z2);
\path (Y1) edge [bend left=0] (Z2);
\node[draw=black,dotted,fit=(Z1) (X1) (Y1), inner sep=0.2cm] (box1) {};
\node[above=2pt of box1, align=center] {1st time interval};
\node[draw=black,dotted,fit=(Z2) (X2) (Y2), inner sep=0.2cm] (box2) {};
\node[above=2pt of box2, align=center] {2nd time interval};
\end{tikzpicture}
         \caption{Reference causal ordering ($Z_t$, $X_t$, $Y_t$), where treatment precedes the outcome within time intervals}
         \vspace{.25cm}
         \label{fig:xt_yt}
     \end{subfigure}
        \caption{Directed acyclic graphs (DAGs) representing the two reference causal orderings in a study with two time intervals ($T = 2$). $X_t$ denotes treatment, $Z_t$ time-varying confounders, and $Y_t$ the outcome. Subscripts indicate study periods. $Z_1$ additionally includes baseline confounders. Variables in $\bar{X}_T$ are shown in blue, whereas $Y_T$ is shown in red, representing the exposure and outcome of the study, respectively.}
        \label{fig:referrence_orderings}
\end{figure}

To operationalize the difference between these two DAGs, we need to specify the possible values of each variable and the mechanisms that influence them. We define $Y_t$ as an absorbing outcome (e.g., death). We set $Y_t = 0$ if the patient survives time interval $t$, and $Y_t = 1$ if the patient dies during interval $t$ or previously. The state $Y_t = 1$ is absorbing for vital status ($Y_{t-1} = 1$ implies $Y_t = 1$), which justifies the directed edge from $Y_{t-1}$ to $Y_t$. In the within-interval causal ordering ($Z_t$, $X_t$, $Y_t$), treatment up to $X_t$ is a direct cause of $Y_t$. In the ordering ($Z_t$, $Y_t$, $X_t$), the absence of a directed edge from $X_t$ to $Y_t$ reflects the assumption that the lag of the treatment effect exceeds the width of the time intervals.

Treatment $X_t$ is directly affected by treatment received in all preceding time intervals and by time-varying confounders up to $Z_t$. We allow $X_t$ to take a special value $u$ (``unclear") whenever the patient died before time interval $t$, thereby justifying directed edges from $Y_{t-1}$ to $X_t$. This reflects the fact that, after death, it is not well-defined whether the patient received treatment in subsequent intervals. When the patient dies during interval $t$ without having received treatment during that interval, we also have $X_t = u$ in the ordering where $Y_t$ precedes $X_t$. Indeed, the patient may have been prevented from receiving treatment due to death: assuming $X_t = 0$ can induce immortal time bias through misclassification of the treatment strategy. If the patient took the treatment during interval $t$ before dying, we set $X_t = 1$, which does not induce immortal time bias.

In the ordering ($Z_t$, $X_t$, $Y_t$), we assume that death during interval $t$ does not affect treatment received within that interval. In other words, the outcome cannot occur between the beginning of an interval and the scheduled treatment administration within that same interval. As a result, if a patient dies during $t$ without having received treatment in that interval, we assume that they would not have received it during the remainder of the interval had they survived. Accordingly, we set $X_t = 0$ in this case.

In both causal orderings, time-varying confounders up to $Z_t$ are direct causes of $Y_t$. Time varying confounders at time $t$ are directly affected by confounders in all preceding time intervals and by treatment up to $X_{t-1}$. Similarly to $X_t$, $Z_t$ has a special value $u$ when the patient died before $t$, thereby justifying a directed edge from $Y_{t-1}$ to $Z_t$.

Although the value $u$ is useful for defining $X_t$ and $Z_t$, it does not appear in the calculations, as shown in Section~\ref{sec:estimation}. As in \cite{cainWhenStartTreatment2010}, we make the untestable assumption of no unmeasured confounding. We also assume positivity, that is, that alive patients have a positive probability of receiving either treatment or no treatment at each time interval and for every observed covariate and treatment history.

\subsection{Clinical plausibility of within-interval causal orderings} \label{sec:plausibility}

In the within-interval causal ordering ($Z_t$, $Y_t$, $X_t$), we assume that the lag between treatment initiation and its effect exceeds the interval width. For example, this assumption is often plausible when studying the effects of preventive treatments on complications of chronic diseases using one-week or even one-month intervals, as these effects typically emerge only after several weeks or longer. Type 2 diabetes mellitus is one such context, as the effect of glucose-lowering medications on vascular complications and mortality has been shown to emerge only after several years of treatment \parencite{ukpdsIntensiveBloodglucoseControl1998}. In dyslipidemia, the effect of statins on cardiovascular event prevention appears within a few weeks in secondary prevention and within a few months in primary prevention \parencite{ridkerRosuvastatinPreventVascular2008, schwartzEffectsAtorvastatinEarly2001}. Similarly, the protective effect of maintenance antidepressant treatment for preventing relapse of depression appears after approximately one month \parencite{lewisMaintenanceDiscontinuationAntidepressants2021}.

In the causal ordering ($Z_t$, $X_t$, $Y_t$), we assume that the outcome cannot occur between the beginning of an interval and the scheduled treatment administration within that same interval. This assumption would hold exactly if treatment could only be administered at the start of each interval. To our knowledge, no clinical setting satisfies this latter condition, as there is always some variability in the timing of treatment administration. Nevertheless, this condition is not necessary for the assumed causal ordering to be reasonable.

Indeed, one-time treatments administered in patients with a low short-term risk of experiencing the outcome, such as elective surgery, may reasonably approximate this ordering. Elective surgery is performed in clinically stable patients. As a result, death in the days preceding a planned surgery is rare. This causal ordering is particularly useful for high-risk elective procedures, as treatment-related complications may occur during or shortly after the intervention, making it inappropriate to assume a long lag between treatment initiation and its effects. For example, 30-day mortality is approximately 5\% following major procedures such as esophagectomy for cancer, pneumonectomy, and coronary artery bypass grafting \parencite{vankootenPatientRelatedPrognosticFactors2022, grapatsasPneumonectomyPrimaryLung2023, shahianPredictorsLongtermSurvival2012}. Similarly, in interventional neuroradiology, endovascular treatment of unruptured intracranial aneurysms is associated with an approximately 5\% risk of severe complications within 30 days \parencite{algraProceduralClinicalComplications2019}. Therefore, this causal ordering may be particularly relevant in elective surgical settings involving time intervals of approximately one week or one month.

\section{Estimation procedures based on causal ordering} \label{sec:estimation}

\subsection{G-computation} \label{sec:gcomp}

When all confounders are observed, g-computation is an algorithm that always provides a valid formula for estimating the expected outcome under a static deterministic treatment strategy \parencite{robinsNewApproachCausal1986, pearlProbabilisticEvaluationSequential1995}. Accordingly, we consider a treatment strategy ``always treat", in which patients are always treated, and a control strategy ``never treat", in which patients are never treated.

Let $T$ denote the number of time intervals defining the study length, and let $Q_1$ and $Q_0$ denote the expected survival under the treatment and control strategies, respectively. Under the reference causal ordering ($Z_t$, $Y_t$, $X_t$), the g-computation algorithm leads to the following expressions (the procedure is detailed in Supplementary material~1):
\begin{align*}
Q_1 =& \sum\limits_{\bar{z}_{T}} \prod\limits_{k=1}^{T} P(Y_k = 0 | \bar{Z}_{k-1}=\bar{z}_{k-1}, Y_{k-1}=0, \bar{X}_{k-1}=1)\\
    &\times \prod\limits_{k=1}^{T-1} P(Z_{k}=z_{k} | \bar{Z}_{k-1}=\bar{z}_{k-1}, Y_{k}=0,\bar{X}_{k-1}=1)\\
Q_0 =& \sum\limits_{\bar{z}_{T}} \prod\limits_{k=1}^{T} P(Y_k = 0 | \bar{Z}_{k-1}=\bar{z}_{k-1}, Y_{k-1}=0, \bar{X}_{k-1}=0)\\
    &\times \prod\limits_{k=1}^{T-1} P(Z_{k}=z_{k} | \bar{Z}_{k-1}=\bar{z}_{k-1}, Y_{k}=0,\bar{X}_{k-1}=0)
\end{align*}

The average treatment effect can then be computed as the difference between $Q_1$ and $Q_0$, or more generally as any contrast between $Q_1$ and $Q_0$. As each term of the formulas involving $Z_t$ is also conditioned on $Y_{t-1} = 0$, the case $Z_t=u$ never appears in the calculations. Similarly, each term involving $X_t$ is conditioned on $Y_t = 0$, so that $X_t = u$ never appears in the formula.

Under the reference causal ordering ($Z_t$, $X_t$, $Y_t$), the g-computation algorithm leads to the following expressions (due to this ordering, $Z=u$ and $X=u$ also never appear):
\begin{align*}
Q_1 =& \sum\limits_{\bar{z}_{T}} \prod\limits_{k=1}^{T} P(Y_k = 0 | \bar{Z}_{k}=\bar{z}_{k}, Y_{k-1}=0, \bar{X}_{k}=1)\\
    &\times \prod\limits_{k=1}^{T} P(Z_{k}=z_{k} | \bar{Z}_{k-1}=\bar{z}_{k-1}, Y_{k-1}=0,\bar{X}_{k-1}=1)\\
Q_0 =& \sum\limits_{\bar{z}_{T}} \prod\limits_{k=1}^{T} P(Y_k = 0 | \bar{Z}_{k}=\bar{z}_{k}, Y_{k-1}=0, \bar{X}_{k}=0)\\
    &\times \prod\limits_{k=1}^{T} P(Z_{k}=z_{k} | \bar{Z}_{k-1}=\bar{z}_{k-1}, Y_{k-1}=0,\bar{X}_{k-1}=0)
\end{align*}

\subsection{Cloning-censoring-weighting} \label{sec:ccw}

Cloning-censoring-weighting (CCW) is a weighting-based method that can accommodate dynamic and stochastic treatment strategies, such as those involving grace periods \parencite{cainWhenStartTreatment2010, webster-clarkDemystifyingCloneCensorWeightingStudying2025}. CCW can also be used for deterministic static treatment strategies \parencite{gaberDeMystifyingCloneCensorWeightMethod2024}.

When comparing the strategies ``always treat" and ``never treat", each patient is represented by two clones: one per strategy. A clone is censored at the first time interval in which the treatment received by the patient is not compatible with the strategy assigned to the clone. Weighting of clones corrects the selection bias induced by artificial censoring. The (unstabilized) weight $W_t$ at interval $t$ is estimated based on the observed treatment and covariate history of patients who survive until the end of every interval $k \leq t$:
\begin{align} \label{eq:weights}
\hat{W}_t = \prod^t_{k=1} \frac{1}{\hat{P}(x_k | \bar{x}_{k-1}, \bar{z}_{k}, \bar{Y}_{k} = 0)}
\end{align}
\color{black}
As with g-computation, $X_t=u$ and $Z_t=u$ never appear because all terms are conditioned on survival.

Finally, $Q_1$ and $Q_0$ are estimated using the clones assigned to the ``always treat" and ``never treat" strategies, respectively. Let $S$ denote the strategy assigned to each clone ($S=1$ for ``always treat" and $S=0$ for ``never treat"). Estimates of $Q_1$ and $Q_0$ are obtained as the products of conditional survival probabilities at each $t$ from the clone data weighted by $W_{t-1}$ (survival during the first interval is estimated without weighting, as artificial censoring cannot occur before the first interval). Here, $C_t=0$ indicates that censoring has not occurred by the end of time interval $t$:
\begin{align*}
\hat{Q}_1 = \prod^T_{t=1} \hat{P}(Y_t = 0 | Y_{t-1}=0, C_{t-1}=0, S=1)\\
\hat{Q}_0 = \prod^T_{t=1} \hat{P}(Y_t = 0 | Y_{t-1}=0, C_{t-1}=0, S=0)
\end{align*}
\color{black}

The validity of CCW for estimating survival under a treatment strategy
$\bar{x}_T$ relies in particular on the following conditional exchangeability assumption~\parencite{cainWhenStartTreatment2010}:
\begin{align} \label{eq:cain}
Y^{\bar{x}_T}_{t}~\indep~X_k \mid \bar{Z}_{k}, \bar{X}_{k-1} = 0, \bar{Y}_{k} = 0 && \text{for all $t, k \in \{1, \dots, T\}$}
\end{align}
where $Y^{\bar{x}_T}_{t}$ is the vital status of the patient during interval $t$ under a hypothetical intervention making them follow strategy~$\bar{x}_T$ ($Y^{\bar{x}_T}_{t}$ is a counterfactual variable). In Supplementary material~2, we use ancestral multi-world networks (introduced by \cite{correaCounterfactualGraphicalModels2025} to assess counterfactual independencies using the d-separation criterion) to show that assumption~\ref{eq:cain} holds in the reference causal ordering ($Z_t$, $Y_t$, $X_t$), but does not hold in the ordering ($Z_t$, $X_t$, $Y_t$).

\subsection{Modification of cloning-censoring-weighting} \label{sec:modif_ccw}

In the reference causal ordering ($Z_t$, $X_t$, $Y_t$), the exchangeability assumption~\ref{eq:cain} does not hold because of conditioning on $Y_k$, as shown in Supplementary material~2. However, conditional exchangeability~\ref{eq:cain} can be restored by relabeling the treatment and covariate variables: for any $t = 1, \dots, T$, $X_t$ becomes $X_{t-1}$ and $Z_t$ becomes $Z_{t-1}$. As a result, the index $t$ no longer corresponds to a time interval; instead, we refer to it as a ``causal interval". The relabeling also introduces the causal interval $t=0$, which could be viewed as the study baseline (in particular, no patient can die at $t=0$). Figure~\ref{fig:relabeling} is a modified version of Figure~\ref{fig:xt_yt} that illustrates the relabeling.

\begin{figure}
     \centering
\begin{tikzpicture}
\node[state, text={rgb:red,39;green,82;blue,184}, draw={rgb:red,39;green,82;blue,184}, fill=white] at (-9.30, 10.50) (X1) {$X_{0}$};
\node[state, text={rgb:red,39;green,82;blue,184}, draw={rgb:red,39;green,82;blue,184}, fill=white] at (-6.00, 10.50) (X2) {$X_{1}$};
\node[state, text=black, draw=black, fill=white] at (-9.30, 7.50) (Y1) {$Y_{1}$};
\node[state, text={rgb:red,230;green,26;blue,25}, draw={rgb:red,230;green,26;blue,25}, fill=white] at (-6.00, 7.50) (Y2) {$Y_{2}$};
\node[state, text=black, draw=black, fill=white] at (-11.10, 9.00) (Z1) {$Z_{0}$};
\node[state, text=black, draw=black, fill=white] at (-7.80, 9.00) (Z2) {$Z_{1}$};
\path (Y1) edge [bend left=0] (Y2);
\path (X1) edge [bend left=0] (X2);
\path[line width=2.5] (X1) edge [bend left=0] (Y1);
\path (Y1) edge [bend left=-24] (X2);
\path[line width=2.5] (X2) edge [bend left=0] (Y2);
\path (X1) edge [bend left=22] (Y2);
\path (Z1) edge [bend left=0] (X1);
\path (Z1) edge [bend left=0] (Y1);
\path (Z1) edge [bend left=-6] (X2);
\path (Z1) edge [bend left=0] (Y2);
\path (Z2) edge [bend left=0] (X2);
\path (Z2) edge [bend left=0] (Y2);
\path (X1) edge [bend left=0] (Z2);
\path (Z1) edge [bend left=0] (Z2);
\path (Y1) edge [bend left=0] (Z2);
\node[draw=black,dotted,fit=(Z1) (X1) (Y1), inner sep=0.2cm] (box1) {};
\node[above=2pt of box1, align=center] {1st time interval};
\node[draw=black,dotted,fit=(Z2) (X2) (Y2), inner sep=0.2cm] (box2) {};
\node[above=2pt of box2, align=center] {2nd time interval};
\end{tikzpicture}
        \caption{Relabeling the variables to restore the exchangeability assumption of cloning-censoring-weighting in the reference causal ordering in which treatment precedes the outcome within each time interval: for any time interval $t$, $X_t$ become $X_{t-1}$ and $Z_t$ become $Z_{t-1}$.}
        \label{fig:relabeling}
\end{figure}

This relabeling leads to two modifications when implementing CCW. First, when estimating the conditional survival probability for a given time interval from the weighted clone data, clones may be censored if they deviate from their assigned strategy during that interval. Thus, censoring is effectively treated as occurring at the beginning of each time interval rather than at its end. Second, weights are modified: they are computed starting from $t=0$, the product in equation~\ref{eq:weights} starts from $k=0$, and the probability of treatment at a given time interval is conditioned on survival through the previous interval rather than on survival through the interval in which the probability of treatment is evaluated. Because the CCW procedure itself is unchanged and assumption~\ref{eq:cain} is restored, the validity of this version of CCW follows directly from the proof provided by \cite{cainWhenStartTreatment2010}.

\subsection{Simulations}
We simulated data compatible with the DAGs of the two reference within-interval causal orderings depicted in Figure~\ref{fig:referrence_orderings}. For each ordering, we estimated the bias of three estimators of the average treatment effect (ATE) on the difference scale: nonparametric CCW \parencite{cainWhenStartTreatment2010}, nonparametric g-computation, and a nonparametric version of the modified CCW introduced in Section~\ref{sec:modif_ccw}. The standard and modified CCW procedures were applied unchanged under both orderings, whereas the g-computation estimand was adapted to the assumed within-interval structure (Equation~\ref{eq:estimand_1} for the ordering $(Z_t, Y_t, X_t)$ and Equation~\ref{eq:estimand_2} for the ordering $(Z_t, X_t, Y_t)$). The censoring and death hazards were estimated with saturated logistic regressions. The data-generating process and the estimation procedures are detailed in Supplementary material~3. We simulated $1{,}000$ datasets of $1{,}000$ patients each, with a follow-up partitioned into two time intervals ($T = 2$), a setting sufficient to reproduce both time-varying confounding and risk of immortal time bias.

\begin{figure}[htbp]
\centering
\begin{subfigure}{0.48\textwidth}
  \centering
  \includegraphics[width=0.88\linewidth]{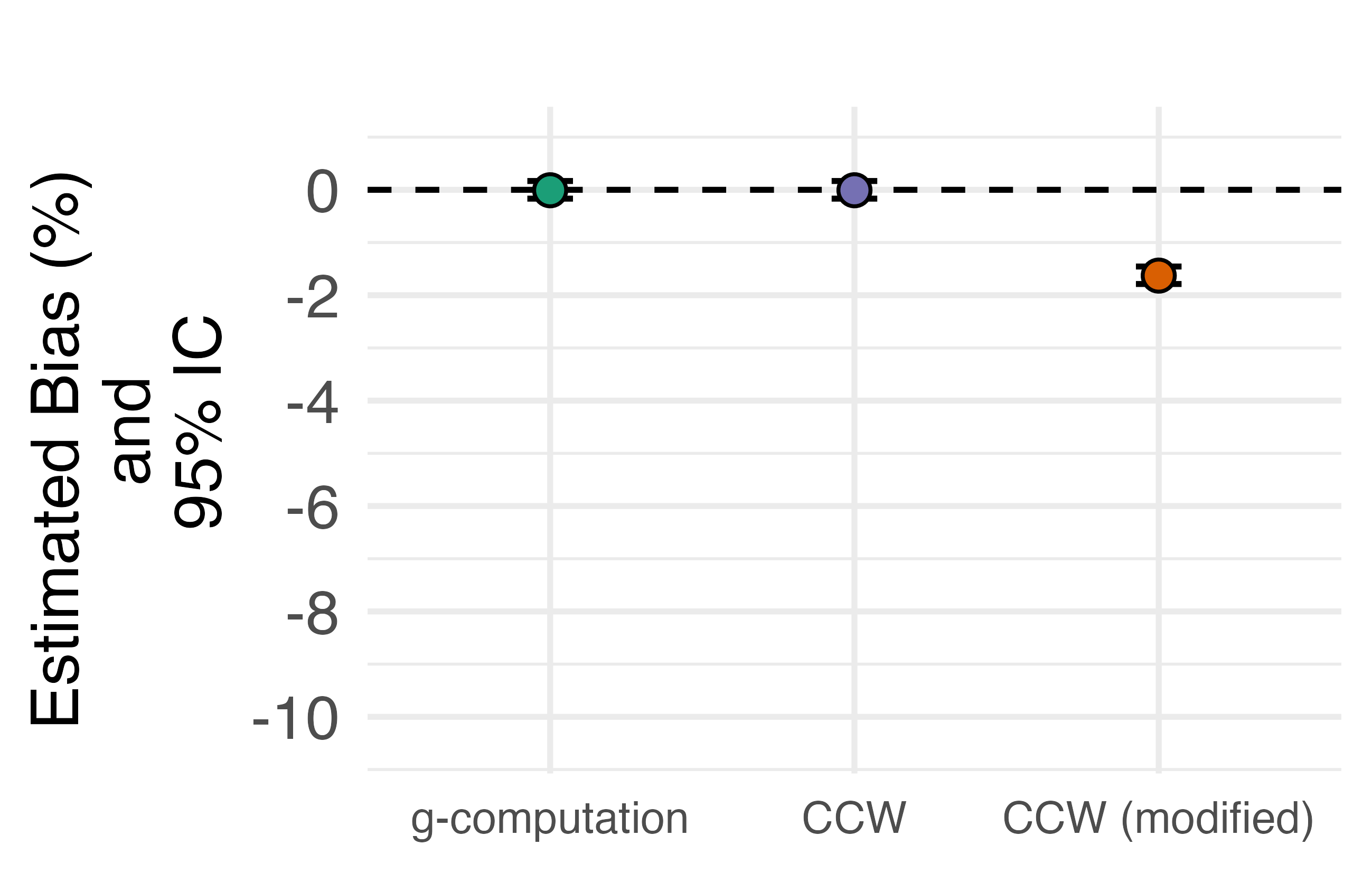}
  \caption{Simulation scenario under the reference causal ordering $(Z_t, Y_t, X_t)$}
  \label{fig:bias_cas_1}
\end{subfigure}
\hfill
\begin{subfigure}{0.48\textwidth}
  \centering
  \includegraphics[width=0.88\linewidth]{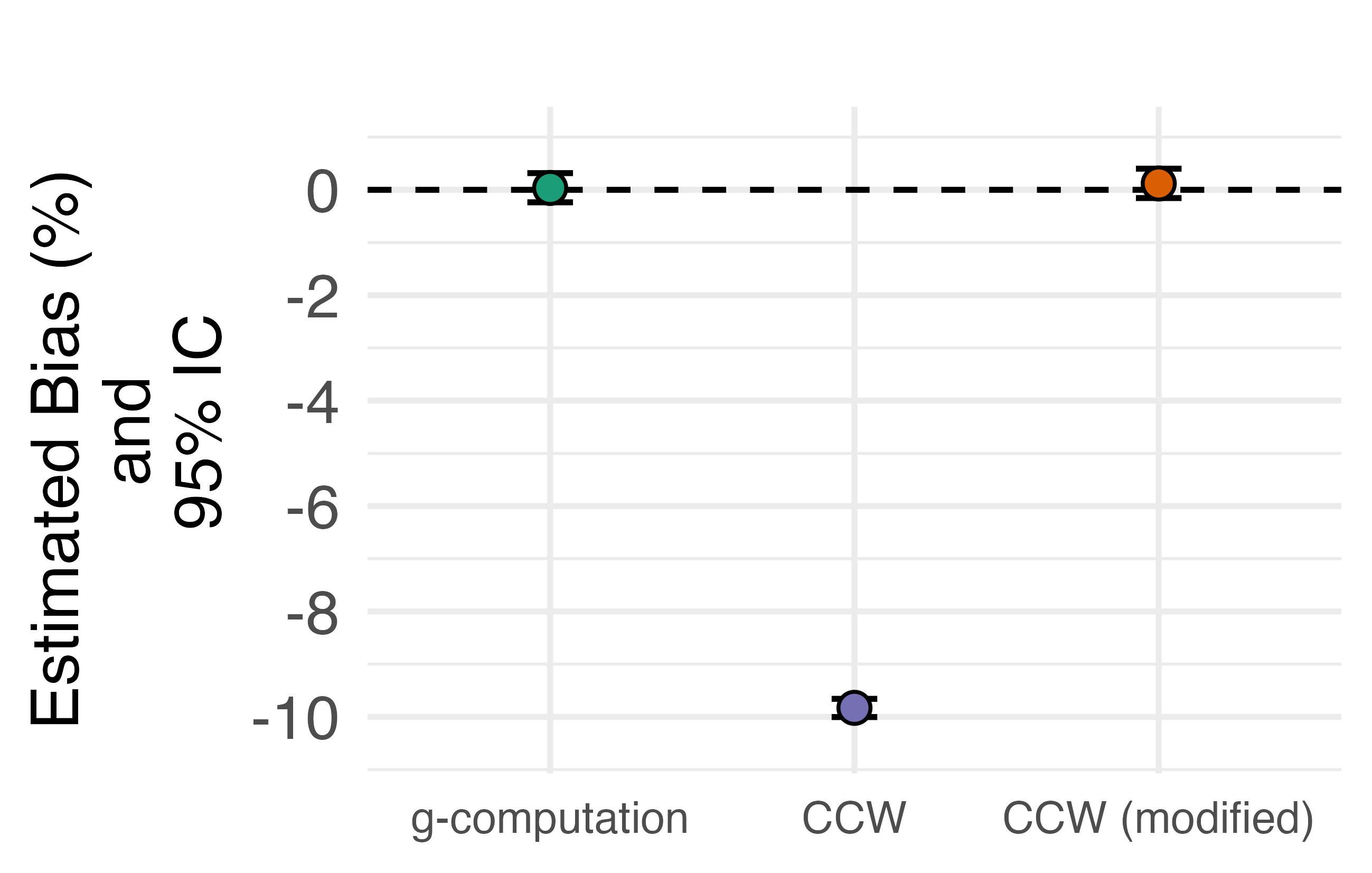}
  \caption{Simulation scenario under the reference causal ordering $(Z_t, X_t, Y_t)$}
  \label{fig:bias_cas_2}
\end{subfigure}
\caption{Simulation results assessing the bias of the nonparametric standard cloning-censoring-weighting (CCW), nonparametric g-computation, and the nonparametric modified CCW (Section~\ref{sec:modif_ccw}). Comparison of ``always treat'' versus ``never treat'' in simple scenarios with two time intervals ($T = 2$). 1,000 simulated datasets of 1,000 patients each. CI: confidence interval.}
\label{fig:simu_ccw}
\end{figure}

For the ordering $(Z_t, Y_t, X_t)$, the confidence intervals for the bias of the standard CCW and of g-computation contained zero, indicating no detectable bias, whereas that of the modified CCW did not (Figure~\ref{fig:bias_cas_1}). With a true increase in survival of $12.5\%$ at the end of the second interval in the ``always treat'' group, the estimated bias was $-0.008\%$ ($95\%$ CI: $-0.17\%$ to $0.17\%$) for the standard CCW, $-0.008\%$ ($95\%$ CI: $-0.17\%$ to $0.13\%$) for g-computation, and $-1.63\%$ ($95\%$ CI: $-1.79\%$ to $-1.46\%$) for the modified CCW.
By contrast, for the ordering $(Z_t, X_t, Y_t)$, the confidence interval for the bias of the standard CCW did not contain zero, whereas those of g-computation and of the modified CCW did (Figure~\ref{fig:bias_cas_2}). With a true increase in survival of $16.1\%$ at the end of the second interval in the ``always treat'' group, the estimated bias was $-9.83\%$ ($95\%$ CI: $-10.0\%$ to $-9.66\%$) for the standard CCW, $0.03\%$ ($95\%$ CI: $-0.24\%$ to $0.32\%$) for g-computation, and $0.12\%$ ($95\%$ CI: $-0.17\%$ to $0.40\%$) for the modified CCW.

Taken together, these results show that the standard CCW is biased precisely when treatment affects the outcome within a time interval (ordering $(Z_t, X_t, Y_t)$), and that the modified CCW restores validity in that case; conversely, the modified CCW is itself biased under the ordering $(Z_t, Y_t, X_t)$. Each CCW variant is thus valid only under the within-interval causal ordering it is designed for, whereas g-computation, with the ordering-appropriate estimand, showed no detectable bias under both. The choice between the standard and modified CCW therefore depends on the assumed within-interval causal ordering. The simulation code is available as an online supplementary resource on \href{https://github.com/Harold229/gcomp-ccw-simulations}{GitHub}.

\subsection{Alternative causal orderings} \label{sec:alternative}

In addition to the reference within-interval causal orderings ($Z_t$, $Y_t$, $X_t$) and ($Z_t$, $X_t$, $Y_t$), four alternative orderings are possible: ($Y_t$, $Z_t$, $X_t$), ($Y_t$, $X_t$, $Z_t$), ($X_t$, $Z_t$, $Y_t$), and ($X_t$, $Y_t$, $Z_t$). In Supplementary material~4, we show that estimation under these alternative causal orderings can be performed in two steps:
\begin{enumerate}
\item If $Z_t$ follows $X_t$ in the within-interval causal ordering, relabel $Z_t$ as $Z_{t+1}$ (and baseline confounders as $Z_1$).
\item The estimation procedure corresponding to the reference causal ordering with the same relative ordering of $X_t$ and $Y_t$ can then be applied.
\end{enumerate}
For example, for the causal ordering ($X_t$, $Z_t$, $Y_t$), $Z_t$ must be relabeled as $Z_{t+1}$. The same g-computation and CCW procedures used for the reference causal ordering ($Z_t$, $X_t$, $Y_t$) can then be applied.

\section{Practical considerations with time partitioning} \label{sec:practical_considerations}

\subsection{Excessively wide time intervals} \label{sec:wide_intervals}

Widening the time intervals can result in causal cycles between variables. For example, with one-year time intervals, oral antidiabetic treatment may affect glycated hemoglobin levels, which may in turn influence treatment decisions before the end of the same interval. In Supplementary material~5, we show that any within-interval causal cycle involving treatment and either the outcome or the covariates leads to non-identifiability of the treatment effect. Non-identifiability means that different causal models with different causal effects can generate the same observed data distribution. Consequently, no statistical procedure can reliably estimate the treatment effect from the observed data alone without additional assumptions. Conversely, we show that causal cycles involving only the outcome and covariates neither lead to non-identifiability nor require modification of the g-computation formula or the CCW procedure.

Widening the time intervals can also create difficulties in defining the causal effect of interest. Consider a six-month study evaluating the effect of an oral antidiabetic treatment on glycated hemoglobin ($Y$). Suppose that follow-up is initially divided into two three-month intervals, with $X_1 = 1$ indicating treatment during the first three months and $X_2 = 1$ indicating treatment during the last three months. Perfect adherence is assumed, and untreated individuals have $X_1 = X_2 = 0$.

If the time intervals are merged into a single six-month period, treatment can be summarized by the variable $X = X_1 + X_2$, representing the total number of three-month intervals during which treatment was received. Although $X$ preserves information on the cumulative amount of treatment received, it no longer captures its timing. This loss of temporal information becomes problematic when estimating the effect of setting $X = 1$ for the entire population, that is, the counterfactual mean $\mathbb{E}[Y^{X=1}]$: individuals with $X = 1$ may have received treatment only during the first three months $(X_1 = 1, X_2 = 0)$ or only during the last three months $(X_1 = 0, X_2 = 1)$. These two treatment strategies are likely to have different effects on glycated hemoglobin measured at the end of follow-up, because treatment administered closer to the outcome assessment would generally have a greater impact. As a result, the value $\mathbb{E}[Y^{X=1}]$ is not unique, and we say that the treatment effect is not well-defined.

By contrast, the effects of setting treatment to $X = 0$ or $X = 2$ are well-defined, because each corresponds to a single treatment strategy: no treatment throughout follow-up and continuous treatment throughout follow-up, respectively. \cite{hernanDoesWaterKill2016} provides a detailed discussion on treatment definition in causal inference.

\subsection{Excessively narrow time intervals}

While excessively wide time intervals result in causal issues (non-identifiability and ill-defined treatment strategies), excessively short time intervals lead to statistical issues.

The number of variables in the causal structure increases linearly with the number of time intervals (for each interval $t$, there are variables $Z_t$, $Y_t$, and $X_t$). However, the number of possible covariate histories $\bar{z}_t$ increases exponentially with $t$: if $Z_t$ is a categorical variable with $k$ categories, there are up to $k^t$ possible sequences $\bar{z}_t$. As this covariate history is accounted for in g-computation and CCW, the number of parameters required in non-parametric implementations of these methods grows at least exponentially with the number of time intervals.

Several simulation studies have shown that this increase in the number of parameters due to overly fine time partitioning leads to increased estimator variance \parencite{adamsImpactTimeSeries2020, sofryginTargetedLearningDaily2019, ferreiraguerraImpactDiscretizationTimeline2020}. 

To mitigate this variance problem, it is possible to reduce the number of parameters to be estimated by introducing additional causal or statistical assumptions. An example of such a causal assumption would be to assume that $Z_t$ affects treatment or outcome only within the next $l$ intervals, with $l$ chosen based on medical considerations. Common statistical assumptions are, for example, to assume no interactions between covariates and treatment or to assume proportional hazards for death or treatment initiation.

\section{Discussion}

In this study, we highlighted the challenges of time partitioning in target trial emulations in which the treatment strategy is not fully determined at the start of follow-up. We then discussed which within-interval causal orderings are plausible in different clinical settings. For each causal ordering, we presented the corresponding g-computation formula, showed that the standard CCW approach is invalid when treatment affects the outcome within time intervals, and proposed a modified version of CCW that is valid under these conditions. Finally, we analyzed the consequences of choosing time intervals that are either excessively wide or excessively narrow, thereby formally establishing the need for time partitioning and providing practical guidance for selecting an appropriate partition based on the clinical setting.

In target trial emulation, time is often partitioned into daily intervals \parencite{maringeReflectionModernMethods2020, beydonLowDoseAspirinCardiovascular2026, urnerVenovenousExtracorporealMembrane2022}. This practice may partly reflect the temporal resolution of the available data. As argued throughout this study, however, the choice of time intervals should instead be guided by the clinical context. On the one hand, target trial emulations in intensive care using daily intervals may lead to non-identifiable treatment effects because covariates can influence treatment decisions within hours, while treatment may in turn affect these covariates within the same interval \parencite{urnerVenovenousExtracorporealMembrane2022}. On the other hand, daily intervals in settings such as elective surgery or the treatment of chronic conditions may unnecessarily increase estimator variance or require strong causal or statistical assumptions to make estimation feasible \parencite{maringeReflectionModernMethods2020, beydonLowDoseAspirinCardiovascular2026}.

We illustrated the challenges of time partitioning using the simplified setting of static deterministic treatment strategies, although methods such as CCW can accommodate more complex strategies, including those with grace periods or treatment decisions based on time-varying covariates. Likewise, g-computation can be extended to handle conditional and stochastic treatment strategies \parencite{correaCalculusStochasticInterventions2020, wanisGracePeriodsComparative2024}, but our focus here was on the challenges posed by time partitioning.

Similarly, we assumed that there was no unmeasured confounding. However, under some causal structures, treatment effects remain identifiable despite unmeasured confounding \parencite{pearlProbabilisticEvaluationSequential1995, shpitserIdentificationConditionalInterventional2006}.

Loss to follow-up was not considered in this study. Inverse probability weighting can account for this type of censoring by estimating the counterfactual outcome under a treatment strategy in which no individuals are lost to follow-up \parencite{robinsMarginalStructuralModels2000, hernanCausalInferenceWhat2020}. Alternatively, loss to follow-up can be handled as a missing data problem \parencite{mohanGraphicalModelsProcessing2021}.

In conclusion, this study provides guidance for partitioning time when designing a target trial emulation, including the choice of time interval length, the assumed within-interval causal structure, and the estimation procedure. We showed that the standard CCW approach is invalid whenever treatment influences the outcome within a time interval, and proposed a modified procedure that restores its validity in this setting. Because the appropriate estimation procedure depends on the assumed causal structure, we believe that reporting the underlying DAG should become a standard component of reporting guidelines for target trial emulation studies \parencite{cashinTransparentReportingObservational2025}.

\section*{Acknowledgements}

The authors are grateful to Alexis F. Guédon, Delphine Paneau, and Thomas Samaille for sharing their expertise in their respective medical specialties.

\printbibliography

\appendix

\clearpage

\etocsettocstyle{\section*{Supplementary Material}}{}
\localtableofcontents

\subsection{Supplementary material~1: Proofs for the estimands} \label{supplements:estimand}

\subsubsection{Reference within-interval causal ordering $(Z_t, Y_t, X_t)$} \label{sup_gcomp_yx}

We first define the queries $Q_1$ and $Q_0$, that is, mathematical formalizations of the effects of treatment strategies ``always treat" and ``never treat" on survival without meeting the absorbing outcome by the end of the study. We cannot just define $Q_1 = P(Y_T=0|\bar{X}_T =  1)$ and $Q_0 = P(Y_T=0|\bar{X}_T =  0)$, because it implies that after death, patients are treated differently depending on their assigned strategy, which is false, as there is no intervention after death. We rather use functions $g_1$ and $g_0$, that describe the daily treatment decision, so that no intervention is applied after death:
\begin{align*}
&g_1(Y_t) = \begin{cases}
			1 & \text{if $Y_t = 0$}\\
			X_t & \text{if $Y_t = 1$}
		\end{cases}\\
&g_0(Y_t) = \begin{cases}
			0 & \text{if $Y_t = 0$}\\
			X_t & \text{if $Y_t = 1$}
		\end{cases}
\end{align*}
where $X_t$ is the value of the treatment a time $t$ observed in the absence of intervention. We then define $Q_1$ and $Q_0$:
\begin{align*}
&Q_1 = P(Y_T = 0 | do(\bar{X}_T = g_1(\bar{Y}_T))) \\
&Q_0 = P(Y_T = 0 | do(\bar{X}_T = g_0(\bar{Y}_T)))
\end{align*}
where $g_1(\bar{Y}_k) = (g_1(Y_1), \dots, g_1(Y_k))$ and $g_0(\bar{Y}_k) = (g_0(Y_1), \dots, g_0(Y_k))$.


In this section, we provide the formula of the estimand for the query $Q_0 = P(Y_{T}=0 | do(\bar{X}_{T} = g_0(\bar{Y}_t))$ where $T$ is any strictly positive integer and the causal graph has causal ordering $(Z_t,Y_t,X_t)$ such as Figure~\ref{fig:yt_xt} extended to $T$ periods. The proof for the estimand in the treated group is similar.

\begin{align*}
Q_0 =& P(Y_{T} = 0 | do(\bar{X}_{T} = g_0(\bar{Y}_T)))\\
    =& P(Y_{T} = 0 | do(\bar{X}_{T-1} = g_0(\bar{Y}_{T-1}))) & \text{Rule 3 of do-calculus}\\
    =& \sum\limits_{\bar{z}_{T-1},\bar{y}_{T-1}} P(Y_T = 0 | \bar{Z}_{T-1}=\bar{z}_{T-1}, \bar{Y}_{T-1}=\bar{y}_{T-1}, \bar{X}_{T-1}=g_0(\bar{Y}_{T-1})) &\text{Sequential backdoor}\\
    &\times \prod\limits_{k=1}^{T-1} P(\bar{Z}_{k}=\bar{z}_{k}, \bar{Y}_{k}=\bar{y}_{k} | \bar{Z}_{k-1}=\bar{z}_{k-1}, \bar{Y}_{k-1}=\bar{y}_{k-1},\bar{X}_{k-1}=g_0(\bar{Y}_{k-1}))\\
    =& \sum\limits_{\bar{z}_{T-1}} P(Y_T = 0 | \bar{Z}_{T-1}=\bar{z}_{T-1}, Y_{T-1}=0, \bar{X}_{T-1}=g_0(\bar{Y}_{T-1})) &\forall t\leq T,~ P(Y_T=0|Y_t=1)=0\\
    &\times \prod\limits_{k=1}^{T-1} P(\bar{Z}_{k}=\bar{z}_{k}, \bar{Y}_{k}=0 | \bar{Z}_{k-1}=\bar{z}_{k-1}, Y_{k-1}=0,\bar{X}_{k-1}=g_0(\bar{Y}_{k-1}))\\
    =& \sum\limits_{\bar{z}_{T-1}} P(Y_T = 0 | \bar{Z}_{T-1}=\bar{z}_{T-1}, Y_{T-1}=0, \bar{X}_{T-1}=0) &g_0(0)=0\\
    &\times \prod\limits_{k=1}^{T-1} P(\bar{Z}_{k}=\bar{z}_{k}, \bar{Y}_{k}=0 | \bar{Z}_{k-1}=\bar{z}_{k-1}, Y_{k-1}=0,\bar{X}_{k-1}=0)\\
    =& \sum\limits_{\bar{z}_{T-1}} P(Y_T = 0 | \bar{Z}_{T-1}=\bar{z}_{T-1}, Y_{T-1}=0, \bar{X}_{T-1}=0) &\text{Trivial simplifications}\\
    &\times \prod\limits_{k=1}^{T-1} P(Z_{k}=z_{k}, Y_{k}=0 | \bar{Z}_{k-1}=\bar{z}_{k-1}, Y_{k-1}=0,\bar{X}_{k-1}=0)\\
    =& \sum\limits_{\bar{z}_{T-1}} P(Y_T = 0 | \bar{Z}_{T-1}=\bar{z}_{T-1}, Y_{T-1}=0, \bar{X}_{T-1}=0) &\text{Chain rule}\\
    &\times \prod\limits_{k=1}^{T-1} P(Z_{k}=z_{k} | \bar{Z}_{k-1}=\bar{z}_{k-1}, Y_{k}=0,\bar{X}_{k-1}=0)\\
    &\times \prod\limits_{k=1}^{T-1} P(Y_{k}=0 | \bar{Z}_{k-1}=\bar{z}_{k-1}, Y_{k-1}=0,\bar{X}_{k-1}=0)\\
    =& \sum\limits_{\bar{z}_{T}} \prod\limits_{k=1}^{T} P(Y_k = 0 | \bar{Z}_{k-1}=\bar{z}_{k-1}, Y_{k-1}=0, \bar{X}_{k-1}=0) &\text{Trivial simplifications}\\
    &\times \prod\limits_{k=1}^{T-1} P(Z_{k}=z_{k} | \bar{Z}_{k-1}=\bar{z}_{k-1}, Y_{k}=0,\bar{X}_{k-1}=0)
\end{align*}

The third rule of do-calculus~\cite{pearlCausalityModelsReasoning2009} can be applied as $Y_T$ and $X_T$ are d-separated given $\bar{X}_{T-1})$ in the graph in which arrows going in $\bar{X}_{T}$ are removed. Indeed, in the graph $X_T$ only has in-going edges, thus once those are removed, $X_T$ is isolated.

The sequential backdoor~\cite{pearlCausalityModelsReasoning2009} can be applied as for all $k<T$, $\bar{Z}_k\cup\bar{Y}_k$ are non-descendants of $X_k$ and $Y_T$ and $X_k$ are d-separated given $\bar{X}_{k-1}\cup\bar{Z}_k\cup\bar{Y}_k$ in the graph in which arrows going in $X_{k+1}\cup\cdots\cup X_{T-1}$ are removed as well as those going out of $X_k$. Indeed, every path between $X_k$ and $Y_T$ must have an edge going in $X_k$ as all edges going out of $X_k$ have been removed. Since all parents of $X_k$ are conditioned on, the path is necessarily blocked.

In conclusion, the estimand is:
\begin{align*}
Q_0 &= P(Y_{T} = 0 | do(\bar{X}_T = g_0(\bar{Y}_T)))\\
    =& \sum\limits_{\bar{z}_{T}} \prod\limits_{k=1}^{T} P(Y_k = 0 | \bar{Z}_{k-1}=\bar{z}_{k-1}, Y_{k-1}=0, \bar{X}_{k-1}=0)\\
    &\times \prod\limits_{k=1}^{T-1} P(Z_{k}=z_{k} | \bar{Z}_{k-1}=\bar{z}_{k-1}, Y_{k}=0,\bar{X}_{k-1}=0)
\end{align*}
and a parallel derivation gives:
\begin{align*}
Q_1 &= P(Y_{T} = 0 | do(\bar{X}_T = g_1(\bar{Y}_T)))\\
    =& \sum\limits_{\bar{z}_{T}} \prod\limits_{k=1}^{T} P(Y_k = 0 | \bar{Z}_{k-1}=\bar{z}_{k-1}, Y_{k-1}=0, \bar{X}_{k-1}=1)\\
    &\times \prod\limits_{k=1}^{T-1} P(Z_{k}=z_{k} | \bar{Z}_{k-1}=\bar{z}_{k-1}, Y_{k}=0,\bar{X}_{k-1}=1)
\end{align*}

\subsubsection{Reference within-interval causal ordering $(Z_t, X_t, Y_t)$} \label{sup_gcomp_xy}

We define the queries $Q_1$ and $Q_0$ using functions $g_1$ and $g_0$ to describe the daily treatment decision, so that no intervention is applied after death (contrary to the other causal ordering, we can intervene on treatment during time interval $t$ even if death occurs during this interval, as the outcome does not influence treatment within time intervals):
\begin{align*}
&g_1(Y_{t-1}) = \begin{cases}
			1 & \text{if $Y_{t-1} = 0$}\\
			X_t & \text{if $Y_{t-1} = 1$}
		\end{cases}\\
&g_0(Y_{t-1}) = \begin{cases}
			0 & \text{if $Y_{t-1} = 0$}\\
			X_t & \text{if $Y_{t-1} = 1$}
		\end{cases}
\end{align*}
We then define $Q_1$ and $Q_0$ as in the other reference causal order:
\begin{align*}
&Q_1 = P(Y_T = 0 | do(\bar{X}_T = g_1(\bar{Y}_{T-1}))) \\
&Q_0 = P(Y_T = 0 | do(\bar{X}_T = g_0(\bar{Y}_{T-1})))
\end{align*}
where $g_1(\bar{Y}_{t-1}) = (1, g_1(Y_1), \dots, g_1(Y_{t-1}))$ and $g_0(\bar{Y}_{t-1}) = (0, g_0(Y_1), \dots, g_0(Y_{t-1}))$.

In this section, we provide the formula of the estimand for the query $Q_0 = P(Y_{T}=0 | do(\bar{X}_{T} = g_0(\bar{Y}_t))$ where $T$ is any strictly positive integer and the causal graph has causal ordering $(Z_t,X_t,Y_t)$ such as Figure~\ref{fig:xt_yt} extended to $T$ periods. The proof for the estimand in the treated group is similar.

\begin{align*}
Q_0 =& P(Y_{T} = 0 | do(\bar{X}_{T} = g_0(\bar{Y}_T)))\\
    =& \sum\limits_{\bar{z}_{T}\bar{y}_{T-1}} P(Y_T = 0 | \bar{Z}_{T}=\bar{z}_{T}, \bar{Y}_{T-1}=\bar{y}_{T-1}, \bar{X}_{T}=g_0(\bar{Y}_{T-1})) &\text{Sequential backdoor}\\
    &\times \prod\limits_{k=1}^{T} P(\bar{Z}_{k}=\bar{z}_{k}, \bar{Y}_{k-1}=\bar{y}_{k-1} | \bar{Z}_{k-1}=\bar{z}_{k-1}, \bar{Y}_{k-2}=\bar{y}_{k-2},\bar{X}_{k-1}=g_0(\bar{Y}_{k-2}))\\
    =& \sum\limits_{\bar{z}_{T}} P(Y_T = 0 | \bar{Z}_{T}=\bar{z}_{T}, Y_{T-1}=0, \bar{X}_{T}=g_0(\bar{Y}_{T-1})) &\forall t\leq T,~ P(Y_T=0|Y_t=1)=0\\
    &\times \prod\limits_{k=1}^{T} P(\bar{Z}_{k}=\bar{z}_{k}, \bar{Y}_{k-1}=0 | \bar{Z}_{k-1}=\bar{z}_{k-1}, Y_{k-2}=0,\bar{X}_{k-1}=g_0(\bar{Y}_{k-2}))\\
    =& \sum\limits_{\bar{z}_{T}} P(Y_T = 0 | \bar{Z}_{T}=\bar{z}_{T}, Y_{T-1}=0, \bar{X}_{T}=0) &g_0(0)=0\\
    &\times \prod\limits_{k=1}^{T} P(\bar{Z}_{k}=\bar{z}_{k}, \bar{Y}_{k-1}=0 | \bar{Z}_{k-1}=\bar{z}_{k-1}, Y_{k-2}=0,\bar{X}_{k-1}=0)\\
    =& \sum\limits_{\bar{z}_{T}} P(Y_T = 0 | \bar{Z}_{T}=\bar{z}_{T}, Y_{T-1}=0, \bar{X}_{T}=0) &\text{Trivial simplifications}\\
    &\times \prod\limits_{k=1}^{T} P(Z_{k}=z_{k}, Y_{k-1}=0 | \bar{Z}_{k-1}=\bar{z}_{k-1}, Y_{k-2}=0,\bar{X}_{k-1}=0)\\
    =& \sum\limits_{\bar{z}_{T}} P(Y_T = 0 | \bar{Z}_{T}=\bar{z}_{T}, Y_{T-1}=0, \bar{X}_{T}=0) &\text{Chain rule}\\
    &\times \prod\limits_{k=1}^{T} P(Z_{k}=z_{k} | \bar{Z}_{k-1}=\bar{z}_{k-1}, Y_{k-1}=0,\bar{X}_{k-1}=0)\\
    &\times \prod\limits_{k=1}^{T} P(Y_{k-1}=0 | \bar{Z}_{k-1}=\bar{z}_{k-1}, Y_{k-2}=0,\bar{X}_{k-1}=0)\\
    =& \sum\limits_{\bar{z}_{T}} \prod\limits_{k=1}^{T} P(Y_k = 0 | \bar{Z}_{k}=\bar{z}_{k}, Y_{k-1}=0, \bar{X}_{k}=0) &\text{Trivial simplifications}\\
    &\times \prod\limits_{k=1}^{T} P(Z_{k}=z_{k} | \bar{Z}_{k-1}=\bar{z}_{k-1}, Y_{k-1}=0,\bar{X}_{k-1}=0)
\end{align*}

The sequential backdoor~\cite{pearlCausalityModelsReasoning2009} can be applied as for all $k\leq T$, $\bar{Z}_k\cup\bar{Y}_{k-1}$ are non-descendants of $X_k$ and $Y_T$ and $X_k$ are d-separated given $\bar{X}_{k-1}\cup\bar{Z}_k\cup\bar{Y}_{k-1}$ in the graph in which arrows going in $X_{k+1}\cup\cdots\cup X_{T}$ are removed as well as those going out of $X_k$. Indeed, every path between $X_k$ and $Y_T$ must have an edge going in $X_k$ as all edges going out of $X_k$ have been removed. Since all parents of $X_k$ are conditioned on, the path is necessarily blocked.

In conclusion, the estimand is:
\begin{align*}
Q_0 =& P(Y_{T} = 0 | do(\bar{X}_{T} = g_0(\bar{Y}_{T-1})))\\
    =& \sum\limits_{\bar{z}_{T}} \prod\limits_{k=1}^{T} P(Y_k = 0 | \bar{Z}_{k}=\bar{z}_{k}, Y_{k-1}=0, \bar{X}_{k}=0)\\
    &\times \prod\limits_{k=1}^{T} P(Z_{k}=z_{k} | \bar{Z}_{k-1}=\bar{z}_{k-1}, Y_{k-1}=0,\bar{X}_{k-1}=0)
\end{align*}
and a parallel derivation gives:
\begin{align*}
Q_1 &= P(Y_{T} = 0 | do(\bar{X}_{T} = g_1(\bar{Y}_{T-1})))\\
    =& \sum\limits_{\bar{z}_{T}} \prod\limits_{k=1}^{T} P(Y_k = 0 | \bar{Z}_{k}=\bar{z}_{k}, Y_{k-1}=0, \bar{X}_{k}=1)\\
    &\times \prod\limits_{k=1}^{T} P(Z_{k}=z_{k} | \bar{Z}_{k-1}=\bar{z}_{k-1}, Y_{k-1}=0,\bar{X}_{k-1}=1)
\end{align*}

\clearpage

\subsection{Supplementary material~2: Assessment of exchangeability in cloning-censoring-weighting}

\subsubsection{Reference within-interval causal ordering $(Z_t, Y_t, X_t)$}  \label{sup_ccw_yx}

Here, we prove that the following statement holds for any study length $T$ in the reference causal ordering $(Z_t, Y_t, X_t)$. The proof is based on ancestral multi-world networks (AMWNs), developed by \cite{correaCounterfactualGraphicalModels2025}:
\begin{align} \label{eq:supp_cain}
Y^{\bar{x}_T}_{t}~\indep~X_k \mid \bar{Z}_{k}, \bar{X}_{k-1} = 0, \bar{Y}_{k} = 0 && \text{for all $t, k \in \{1, \dots, T\}$}
\end{align}

Let $T$ be a positive integer and let $G$ be an extended version of the DAG in Figure~\ref{fig:yt_xt} (main text) to $T$ time intervals. If $t = 1$ and $k \in \{1, \dots, T\}$, statement~\ref{eq:supp_cain} corresponds to conditional independence between $\lVert Y_{t}^{\bar{x}_{T}} \rVert = Y_1$ ($\lVert .\lVert$ is the exclusion operator of \cite{correaCounterfactualGraphicalModels2025}, which removes interventions that have no effects) and $X_k$. As $Y_1$ is also in the conditioning set, statement~\ref{eq:supp_cain} trivially holds.

Let $t \in \{2, \dots, T\}$ and $k \in \{1, \dots, T\}$. Consider $G'$, the AMWN built from $G$ and the variables in statement~\ref{eq:supp_cain} (Figure~\ref{fig:amwn_yt_xt} illustrates this AMWN for $T=2, k=2, t=2$). $G'$ includes the ancestors of $X_k$, that is, $X_k$ and $\bar{X}_{k-1}, \bar{Z}_{k}, \bar{Y}_{k}$, corresponding to the conditioning set in statement~\ref{eq:supp_cain}. $G'$ also includes the ancestors of $Y^{\bar{x}_{T}}_t = \lVert Y^{\bar{x}_T}_t \rVert= Y^{\bar{x}_{t-1}}_t$, that is $Y^{\bar{x}_{t-1}}_t$ and $\lVert \bar{Z}_t^{\bar{x}_{t-1}} \rVert, \lVert \bar{Y}_{t-1}^{\bar{x}_{t-2}} \rVert$.

\begin{figure}
     \centering
\begin{tikzpicture}
\node[rectangle, text=black, draw=black, fill=white] at (-9.30, 10.50) (X1) {$X_{1}$};
\node[state, opacity=0, text=black, draw=black, fill=white] at (-9.30, 10.50) (Xa) {$X_{a}$};
\node[state, text={rgb:red,0;green,82;blue,64}, draw={rgb:red,0;green,82;blue,64}, fill=white] at (-6.00, 10.50) (X2) {$X_{2}$};
\node[rectangle, text=black, draw=black, fill=white] at (-9.30, 7.50) (Y1) {$Y_{1}$};
\node[rectangle, text=black, draw=black, fill=white] at (-6.00, 7.50) (Y2) {$Y_{2}$};
\node[rectangle, text=black, draw=black, fill=white] at (-11.10, 9.00) (Z1) {$Z_{1}$};
\node[rectangle, text=black, draw=black, fill=white] at (-7.80, 9.00) (Z2) {$Z_{2}$};
\node[state, text={rgb:red,0;green,82;blue,64}, draw={rgb:red,0;green,82;blue,64}, fill=white] at (-6.00, 6.00) (Y2*) {$Y_{2}^{x_1}$};
\node[state, text=black, draw=black, fill=white] at (-7.80, 6.00) (Z2*) {$Z_{2}^{x_1}$};
\path (Y1) edge [bend left=0] (Y2);
\path (X1) edge [bend left=0] (X2);
\path[line width=2.5] (Y1) edge [bend left=0] (X1);
\path (Y1) edge [bend left=-24] (X2);
\path[line width=2.5] (Y2) edge [bend left=0] (X2);
\path (X1) edge [bend left=22] (Y2);
\path (Z1) edge [bend left=0] (X1);
\path (Z1) edge [bend left=0] (Y1);
\path (Z1) edge [bend left=-6] (X2);
\path (Z1) edge [bend left=0] (Y2);
\path (Z2) edge [bend left=0] (X2);
\path (Z2) edge [bend left=0] (Y2);
\path (X1) edge [bend left=0] (Z2);
\path (Z1) edge [bend left=0] (Z2);
\path (Y1) edge [bend left=0] (Z2);
\path (Z1) edge [out=-25, in=140] (Y2*);
\path (Y1) edge [bend left=0] (Y2*);
\path (Z2*) edge [bend left=0] (Y2*);
\path[bidirected] (Y2) edge [bend left=0] (Y2*);
\path[bidirected] (Z2) edge [bend left=0] (Z2*);
\node[draw=black,dotted,fit=(Z1) (X1) (Y1) (Xa), inner sep=0.2cm] (box1) {};
\node[above=2pt of box1, align=center] {1st time interval};
\node[draw=black,dotted,fit=(Z2) (X2) (Y2), inner sep=0.2cm] (box2) {};
\node[above=2pt of box2, align=center] {2nd time interval};
\end{tikzpicture}
         \caption{Ancestral multi-world network for the reference ordering in which the outcome precedes treatment. Squares indicate the conditioning set in the conditional independence statement being assessed, and variables shown in green are those for which conditional independence is being evaluated.}
         \label{fig:amwn_yt_xt}
\end{figure}

In G', we have that:
\begin{itemize}
    \item There is no bidirected edge pointing at $X_k$, as no other counterfactual instance of $X_k$ is present in the AMWN and there is no bidirected edge in $G$ (no unmeasured confounding).
    \item The parents of $X_k$ correspond exactly to the conditioning set in statement~\ref{eq:supp_cain}.
    \item $Y^{\bar{x}_{t-1}}_t$ is not a descendant of $X_k$.
\end{itemize}
As a result, every path from $X_k$ to $Y^{\bar{x}_{t-1}}_t$ with an edge pointing to $X_k$ is blocked (by the conditioning on all the parents of $X_k$ in $G'$ and the absence of bidirected edge pointing to $X_k$). As $Y^{\bar{x}_{t-1}}_t$ is not a descendant of $X_k$, every path from $X_k$ to $Y^{\bar{x}_{t-1}}_t$ with an edge emerging from $X_k$ contains a collider which is a descendant of $X_k$.
This collider and its descendants cannot be in the conditioning set, as this set corresponds exactly to the set of the parents of $X_k$, so the path is blocked.

Using the soundness of d-separation in AMWNs, we conclude that for any study length $T$, as statement~\ref{eq:supp_cain} holds in $G'$ for all $t, k \in \{1, \dots, T\}$, statement~\ref{eq:supp_cain} also holds in all distributions compatible with $G$.

\subsubsection{Reference within-interval causal ordering $(Z_t, X_t, Y_t)$}

Here, we prove that for any study length $T$, there exists $t,k \in \{1, \dots, T\}$ such that statement~\ref{eq:supp_cain} does not hold in the reference causal ordering $(Z_t, Y_t, X_t)$.

Let $T$ be a positive integer and let $G$ be an extended version of the DAG in Figure~\ref{fig:xt_yt} (main text) to $T$ time intervals. Consider $G'$, the AMWN built from $G$ and the variables in statement~\ref{eq:supp_cain} with $t = k \in \{1, \dots, T\}$. $G'$ includes at least $X_k$, $Y_k$, and $Y^{\bar{x}_{T}}_k = \lVert Y^{\bar{x}_T}_k \rVert= Y^{\bar{x}_{k}}_k$, because these variables are part of statement~\ref{eq:supp_cain}. In $G'$, there is a path between $X_k$ and $Y^{\bar{x}_{k}}_k$ which is composed of an edge from $X_k$ to $Y_k$ pointing at $Y_k$ and a bidirected edge between $Y_k$ and $Y^{\bar{x}_{k}}_k$. As $Y_k$ is a collider and as $Y_k$ is in the conditioning set of statement~\ref{eq:supp_cain}, the path is open, and this statement does not hold in $G'$ when $t = k \in \{1, \dots, T\}$. Using the completeness of d-separation in AMWNs, we conclude that for all study length $T$, there exists at least one distribution compatible with $G$ in which statement~\ref{eq:supp_cain} does not hold.

\begin{figure}
     \centering
\begin{tikzpicture}
\node[rectangle, text=black, draw=black, fill=white] at (-9.30, 10.50) (X1) {$X_{1}$};
\node[state, opacity=0, text=black, draw=black, fill=white] at (-9.30, 10.50) (Xa) {$X_{a}$};
\node[state, text={rgb:red,230;green,26;blue,25}, draw={rgb:red,230;green,26;blue,25}, fill=white] at (-6.00, 10.50) (X2) {$X_{2}$};
\node[rectangle, text=black, draw=black, fill=white] at (-9.30, 7.50) (Y1) {$Y_{1}$};
\node[rectangle, text=black, draw=black, fill=white] at (-6.00, 7.50) (Y2) {$Y_{2}$};
\node[rectangle, text=black, draw=black, fill=white] at (-11.10, 9.00) (Z1) {$Z_{1}$};
\node[rectangle, text=black, draw=black, fill=white] at (-7.80, 9.00) (Z2) {$Z_{2}$};
\node[state, text={rgb:red,230;green,26;blue,25}, draw={rgb:red,230;green,26;blue,25}, fill=white] at (-6.00, 6.00) (Y2*) {$Y_{2}^{\bar{x}_2}$};
\node[state, text=black, draw=black, fill=white] at (-7.80, 6.00) (Z2*) {$Z_{2}^{x_1}$};
\node[state, text=black, draw=black, fill=white] at (-9.30, 6.00) (Y1*) {$Y_{1}^{x_1}$};
\path (Y1) edge [bend left=0] (Y2);
\path (X1) edge [bend left=0] (X2);
\path[line width=2.5] (X1) edge [bend left=0] (Y1);
\path (Y1) edge [bend left=-24] (X2);
\path[line width=2.5, color={rgb:red,230;green,26;blue,25}] (X2) edge [bend left=0] (Y2);
\path (X1) edge [bend left=22] (Y2);
\path (Z1) edge [bend left=0] (X1);
\path (Z1) edge [bend left=0] (Y1);
\path (Z1) edge [bend left=-6] (X2);
\path (Z1) edge [bend left=0] (Y2);
\path (Z2) edge [bend left=0] (X2);
\path (Z2) edge [bend left=0] (Y2);
\path (X1) edge [bend left=0] (Z2);
\path (Z1) edge [bend left=0] (Z2);
\path (Y1) edge [bend left=0] (Z2);
\path (Z1) edge [out=-25, in=135] (Y2*);
\path (Y1*) edge [out=42, in=150] (Y2*);
\path (Z2*) edge [bend left=0] (Y2*);
\path[bidirected, color={rgb:red,230;green,26;blue,25}] (Y2) edge [bend left=0] (Y2*);
\path[bidirected] (Z2) edge [bend left=0] (Z2*);
\path[bidirected] (Y1) edge [bend left=0] (Y1*);
\node[draw=black,dotted,fit=(Z1) (X1) (Y1) (Xa), inner sep=0.2cm] (box1) {};
\node[above=2pt of box1, align=center] {1st time interval};
\node[draw=black,dotted,fit=(Z2) (X2) (Y2), inner sep=0.2cm] (box2) {};
\node[above=2pt of box2, align=center] {2nd time interval};
\end{tikzpicture}
         \caption{Ancestral multi-world network for the reference ordering in which treatment precedes the outcome. Squares indicate the conditioning set in the conditional independence statement being assessed, and variables shown in red are those for which conditional independence is being evaluated. The causal path highlighted in red is active.}
         \label{fig:amwn_xt_yt}
\end{figure}

\subsubsection{Relabeling $Z_t$ and $X_t$ in reference causal ordering $(Z_t, X_t, Y_t)$}  \label{sup_ccw_xy}

Here, we assess statement~\ref{eq:supp_cain} in causal ordering $(Z_t, X_t, Y_t)$ after relabeling $X_t$ to $X_{t-1}$ and $Z_t$ to $Z_{t-1}$.

The proof is the same as in the reference causal ordering $(Z_t, Y_t, X_t)$: for study length $T$, we build the AMWN $G'$ from $G$ and the variables in statement~\ref{eq:supp_cain} for arbitrary $t \in {1, \dots, T}$ and $k \in {0, \dots, T-1}$ and note that there is no bidirected edge pointing at $X_k$, that the parents of $X_k$ correspond exactly to the conditioning set in statement~\ref{eq:supp_cain}, and that $Y^{\bar{x}_{t-1}}_t$ is not a descendant of $X_k$. We deduce that all paths between $X_t$ and $Y^{\bar{x}_{t-1}}_t$ are blocked, so that statement~\ref{eq:supp_cain} holds in $G'$ and, as a result, in all distributions compatible with $G$. Figure~\ref{fig:relabel} illustrates the relabeling of $X_t$ and $Z_t$ in an AMWN with $T=2$, $t=2$, and $k=1$.

\begin{figure}
     \centering
\begin{tikzpicture}
\node[rectangle, text=black, draw=black, fill=white] at (-9.30, 10.50) (X1) {$X_{0}$};
\node[state, opacity=0, text=black, draw=black, fill=white] at (-9.30, 10.50) (Xa) {$X_{a}$};
\node[state, text={rgb:red,0;green,82;blue,64}, draw={rgb:red,0;green,82;blue,64}, fill=white] at (-6.00, 10.50) (X2) {$X_{1}$};
\node[rectangle, text=black, draw=black, fill=white] at (-9.30, 7.50) (Y1) {$Y_{1}$};
\node[state, opacity=0, text=black, draw=black, fill=white] at (-9.30, 7.50) (Ya) {$Y_{1}$};
\node[state, text=black, draw=black, fill=white] at (-6.00, 7.50) (Y2) {$Y_{2}$};
\node[rectangle, text=black, draw=black, fill=white] at (-11.10, 9.00) (Z1) {$Z_{0}$};
\node[rectangle, text=black, draw=black, fill=white] at (-7.80, 9.00) (Z2) {$Z_{1}$};
\node[state, text={rgb:red,0;green,82;blue,64}, draw={rgb:red,0;green,82;blue,64}, fill=white] at (-6.00, 6.00) (Y2*) {$Y_{2}^{\bar{x}_1}$};
\node[state, text=black, draw=black, fill=white] at (-7.80, 6.00) (Z2*) {$Z_{1}^{x_0}$};
\node[state, text=black, draw=black, fill=white] at (-9.30, 6.00) (Y1*) {$Y_{1}^{x_0}$};
\path (Y1) edge [bend left=0] (Y2);
\path (X1) edge [bend left=0] (X2);
\path[line width=2.5] (X1) edge [bend left=0] (Y1);
\path (Y1) edge [bend left=-24] (X2);
\path[line width=2.5] (X2) edge [bend left=0] (Y2);
\path (X1) edge [bend left=22] (Y2);
\path (Z1) edge [bend left=0] (X1);
\path (Z1) edge [bend left=0] (Y1);
\path (Z1) edge [bend left=-6] (X2);
\path (Z1) edge [bend left=0] (Y2);
\path (Z2) edge [bend left=0] (X2);
\path (Z2) edge [bend left=0] (Y2);
\path (X1) edge [bend left=0] (Z2);
\path (Z1) edge [bend left=0] (Z2);
\path (Y1) edge [bend left=0] (Z2);
\path (Z1) edge [out=-25, in=135] (Y2*);
\path (Y1*) edge [out=42, in=150] (Y2*);
\path (Z2*) edge [bend left=0] (Y2*);
\path[bidirected] (Y2) edge [bend left=0] (Y2*);
\path[bidirected] (Z2) edge [bend left=0] (Z2*);
\path[bidirected] (Y1) edge [bend left=0] (Y1*);
\node[draw=black,dotted,fit=(Z1) (X1) (Y1) (Xa) (Ya), inner sep=0.2cm] (box1) {};
\node[above=2pt of box1, align=center] {1st time interval};
\node[draw=black,dotted,fit=(Z2) (X2) (Y2), inner sep=0.2cm] (box2) {};
\node[above=2pt of box2, align=center] {2nd time interval};
\end{tikzpicture}
         \caption{Ancestral multi-world network for the reference ordering in which treatment precedes the outcome, after relabeling $X_t$ and $Z_t$. Squares indicate the conditioning set in the conditional independence statement being assessed, and variables shown in green are those for which conditional independence is being evaluated. Relabeling removes the conditioning on $Y_2$, which restores d-separation.}
         \label{fig:relabel}
\end{figure}

\clearpage
\subsection{Supplementary material~3: Simulation methods}

\subsubsection{Data-generating processes}

\paragraph{First scenario: ordering $(Z_t, Y_t, X_t)$ ($X_t\not\rightarrow Y_t$)}
We simulated 1,000 datasets of 1,000 patients compatible with the DAG of the reference within-interval causal ordering $(Z_t, Y_t, X_t)$ (Figure~\ref{fig:yt_xt}), with a time-varying confounder $Z_t$. As in the main text, $Z_1$ also includes baseline confounders.
All variables followed Bernoulli distributions conditional on their parents (i.e., their immediate causes), with probabilities provided in Table~\ref{table:sim_1}.
\begin{table}[ht]
\centering
\caption{Data-generating process for the first simulation scenario (ordering $(Z_t, Y_t, X_t)$)}
\begin{tabular}{ll}
\hline
\textbf{Expression} & \textbf{Probability} \\
\hline
\multicolumn{2}{l}{\textit{Confounder process}} \\
$P(Z_1 = 1)$ & $0.5$ \\
$P(Z_2 = 1 \mid Z_1, X_1, Y_1 = 0)$ & $0.25 + 0.45 \times Z_1 - 0.15 \times X_1$ \\
\hline
\multicolumn{2}{l}{\textit{Outcome process}} \\
$P(Y_1 = 1 \mid Z_1)$ & $0.20 + 0.15 \times Z_1$ \\
$P(Y_2 = 1 \mid Z_1, Z_2, X_1, Y_1 = 0)$ & $0.25 + 0.10 \times Z_1 + 0.15 \times Z_2 - 0.15 \times X_1$ \\
\hline
\multicolumn{2}{l}{\textit{Treatment process}} \\
$P(X_1 = 1 \mid Z_1, Y_1 = 0)$ & $0.30 + 0.40 \times Z_1$ \\
$P(X_2 = 1 \mid Z_1, Z_2, X_1, Y_2 = 0)$ & $0.20 + 0.40 \times Z_1 + 0.10 \times Z_2 + 0.20 \times X_1$ \\
\hline
\end{tabular}
\label{table:sim_1}
\end{table}
Consistent with the ordering $(Z_t, Y_t, X_t)$, treatment $X_2$ has no effect on the outcome within its interval. We do not describe the generation of treatment and time-varying confounders after death (the value $u$ in the main text), as this information is never used during estimation.

\paragraph{Second scenario: ordering $(Z_t, X_t, Y_t)$ ($Y_t\not\rightarrow X_t$)}
We simulated 1,000 datasets of 1,000 patients compatible with the DAG of the reference within-interval causal ordering $(Z_t, X_t, Y_t)$ (Figure~\ref{fig:xt_yt}), with a time-varying confounder $Z_t$. As in the main text, $Z_1$ also includes baseline confounders.
All variables followed Bernoulli distributions conditional on their parents, with probabilities provided in Table~\ref{table:sim_2}.
\begin{table}[ht]
\centering
\caption{Data-generating process for the second simulation scenario (ordering $(Z_t, X_t, Y_t)$)}
\begin{tabular}{ll}
\hline
\textbf{Expression} & \textbf{Probability} \\
\hline
\multicolumn{2}{l}{\textit{Confounder process}} \\
$P(Z_1 = 1)$ & $0.5$ \\
$P(Z_2 = 1 \mid Z_1, X_1, Y_1 = 0)$ & $0.25 + 0.45 \times Z_1 - 0.15 \times X_1$ \\
\hline
\multicolumn{2}{l}{\textit{Treatment process}} \\
$P(X_1 = 1 \mid Z_1)$ & $0.30 + 0.40 \times Z_1$ \\
$P(X_2 = 1 \mid Z_1, Z_2, X_1, Y_1 = 0)$ & $0.20 + 0.40 \times Z_1 + 0.10 \times Z_2 + 0.20 \times X_1$ \\
\hline
\multicolumn{2}{l}{\textit{Outcome process}} \\
$P(Y_1 = 1 \mid Z_1, X_1)$ & $0.20 + 0.15 \times Z_1 - 0.10 \times X_1$ \\
$P(Y_2 = 1 \mid Z_1, Z_2, X_1, X_2, Y_1 = 0)$ & $0.25 + 0.15 \times Z_1 + 0.15 \times Z_2 - 0.05 \times X_1 - 0.05 \times X_2$ \\
\hline
\end{tabular}
\label{table:sim_2}
\end{table}
Consistent with the ordering $(Z_t, X_t, Y_t)$, treatment up to $X_t$ is a direct cause of $Y_t$ within its interval. We do not describe the generation of treatment and time-varying confounders after death (the value $u$ in the main text), as this information is never used during estimation.

\subsubsection{Estimation procedures}

\paragraph{G-computation}
For each scenario, g-computation was applied with the estimand corresponding to its within-interval ordering, derived in Section~\ref{sup_gcomp_yx} for $(Z_t, Y_t, X_t)$ and in Section~\ref{sup_gcomp_xy} for $(Z_t, X_t, Y_t)$. For the first scenario, we used a fully factorized version of the estimand:
\begin{align}
    \label{eq:estimand_1}
    \texttt{ATE} =\ & \sum_{z_1, z_2} P(Z_1 = z_1)\, P(Y_1 = 0 \mid Z_1 = z_1)\, P(Z_2 = z_2 \mid Z_1 = z_1, X_1 = 1, Y_1 = 0) \nonumber \\
    & \qquad \times\, P(Y_2 = 0 \mid Z_1 = z_1, Z_2 = z_2, X_1 = 1, Y_1 = 0) \nonumber \\
    & -\, \sum_{z_1, z_2} P(Z_1 = z_1)\, P(Y_1 = 0 \mid Z_1 = z_1)\, P(Z_2 = z_2 \mid Z_1 = z_1, X_1 = 0, Y_1 = 0) \nonumber \\
    & \qquad \times\, P(Y_2 = 0 \mid Z_1 = z_1, Z_2 = z_2, X_1 = 0, Y_1 = 0)
\end{align}
Unlike the general estimand, which marginalizes over the confounder history only up to $Z_{t-1}$, this fully factorized form makes the summation over $Z_t$ explicit as well. The two expressions are equivalent, but the factorized one is more convenient here: because the outcome $Y_2$ also depends on $Z_2$ (a mediator on the path from $X_1$ to $Y_2$), summing explicitly over $Z_2$ maps the estimand directly onto the data-generating factors and yields its true value in closed form.
For the second scenario, the corresponding estimand is:
\begin{align}
    \label{eq:estimand_2}
    \texttt{ATE} =\ & \sum_{z_1, z_2} P(Z_1 = z_1)\, P(Y_1 = 0 \mid Z_1 = z_1, X_1 = 1)\, P(Z_2 = z_2 \mid Z_1 = z_1, X_1 = 1, Y_1 = 0) \nonumber \\
    & \qquad \times\, P(Y_2 = 0 \mid Z_1 = z_1, Z_2 = z_2, X_1 = 1, X_2 = 1, Y_1 = 0) \nonumber \\
    & -\, \sum_{z_1, z_2} P(Z_1 = z_1)\, P(Y_1 = 0 \mid Z_1 = z_1, X_1 = 0)\, P(Z_2 = z_2 \mid Z_1 = z_1, X_1 = 0, Y_1 = 0) \nonumber \\
    & \qquad \times\, P(Y_2 = 0 \mid Z_1 = z_1, Z_2 = z_2, X_1 = 0, X_2 = 0, Y_1 = 0)
\end{align}
In both scenarios, each conditional probability was estimated using a saturated logistic regression, yielding a nonparametric maximum likelihood (g-computation) estimator for the average treatment effect (ATE), comparing the strategies ``always treat" versus ``never treat" on the difference scale.

\paragraph{Cloning-censoring-weighting}
For both scenarios, we implemented the standard CCW procedure described in Section~\ref{sec:ccw}. Each patient was represented by two clones, one per strategy (``always treat" and ``never treat"), and a clone was censored at the first interval in which the observed treatment was incompatible with its assigned strategy. Survival under each strategy was obtained as the product over intervals of the weighted conditional survival probabilities, and the ATE as the contrast between the two strategies. Two features characterize this standard implementation, which the modification below reverses. First, a clone deviating from its assigned strategy during an interval is censored from the next interval onward, so it still contributes to that interval's death hazard, which is therefore estimated among clones uncensored through the previous interval. Second, the censoring weights are conditioned on survival within each interval. The weights follow \cite{cainWhenStartTreatment2010}, and the treatment and death hazards were estimated nonparametrically with saturated logistic regressions, conditioning on the full covariate and treatment history up to each interval. The same procedure was applied to the datasets of both scenarios.

\paragraph{Modified cloning-censoring-weighting}
To assess the modified CCW procedure introduced in Section~\ref{sec:modif_ccw}, we applied it to the datasets of both scenarios, as for g-computation and standard CCW, so that its validity could be examined both under the ordering $(Z_t, X_t, Y_t)$, for which it is intended, and under the ordering $(Z_t, Y_t, X_t)$. No additional data were generated. The procedure consists of the index relabeling described in the main text: for each interval $t$, $X_t$ becomes $X_{t-1}$ and $Z_t$ becomes $Z_{t-1}$, which introduces the causal interval $t = 0$ (no patient can die at $t = 0$); vital status $Y_t$ is left unchanged.
Under this relabeling, treatment precedes the outcome within each causal interval, which restores the conditional exchangeability assumption of CCW when the data were themselves generated under the ordering $(Z_t, X_t, Y_t)$ (Table~\ref{table:sim_2}). Two modifications follow when implementing the method. First, a clone deviating from its assigned strategy during an interval is censored at the beginning of that interval rather than at its end; accordingly, the death hazard for that interval is estimated among clones still uncensored at its beginning, whereas in the original CCW it is estimated among clones uncensored through the previous interval. Second, the censoring weights are computed from $t = 0$ onward, and the probability of remaining uncensored at a given causal interval is no longer conditioned on survival within that interval. The weights follow \cite{cainWhenStartTreatment2010}, and the treatment and death hazards were estimated with saturated logistic regressions. The ATE was computed using the same approach as for the standard CCW procedure.

\paragraph{Bias assessment}
For each scenario, the true ATE was computed from the corresponding estimand (Eqs.~\ref{eq:estimand_1}--\ref{eq:estimand_2}) at the true parameter values given in Tables~\ref{table:sim_1} and~\ref{table:sim_2}. Each method was applied to $1{,}000$ replicate datasets, and bias was estimated as the average difference between the estimated and true ATE across replicates, reported in percentage points. The $95\%$ confidence interval for the mean bias was obtained by nonparametric bootstrap of the simulation-level estimates ($R = 1{,}000$ resamples, percentile method), separately for each method.

\clearpage

\subsection{Supplementary material 4: Estimation in alternative causal orderings}

In addition to the reference within-interval causal orderings ($Z_t$, $Y_t$, $X_t$) and ($Z_t$, $X_t$, $Y_t$), four alternative orderings are possible: ($Y_t$, $Z_t$, $X_t$), ($Y_t$, $X_t$, $Z_t$), ($X_t$, $Z_t$, $Y_t$), and ($X_t$, $Y_t$, $Z_t$). In this section, we show that estimation under these alternative causal orderings can be performed in two steps:
\begin{enumerate}
\item If $Z_t$ follows $X_t$ in the within-interval causal ordering, relabel $Z_t$ as $Z_{t+1}$ (and baseline confounders as $Z_1$).
\item The estimation procedure corresponding to the reference causal ordering with the same relative ordering of $X_t$ and $Y_t$ can then be applied.
\end{enumerate}
For example, for the causal ordering ($X_t$, $Z_t$, $Y_t$), $Z_t$ must be relabeled as $Z_{t+1}$ (and baseline confounders as $Z_1$). The same g-computation and CCW procedures used for the reference causal ordering ($Z_t$, $X_t$, $Y_t$) can then be applied.

\textbf{Proof:}\\
For the causal ordering ($Y_t$, $X_t$, $Z_t$), relabeling $Z_t$ as $Z_{t+1}$ and baseline confounders as $Z_1$ yields the same DAG as the reference causal ordering ($Z_t$, $Y_t$, $X_t$). Hence, the g-computation and CCW procedures used for the reference causal ordering ($Z_t$, $Y_t$, $X_t$) can be applied to the alternative causal ordering ($Y_t$, $X_t$, $Z_t$) after this relabeling.\\
For the causal ordering ($X_t$, $Y_t$, $Z_t$), the same argument allows the g-computation and CCW procedures used for the reference causal ordering ($Z_t$, $X_t$, $Y_t$) to be applied after relabeling $Z_t$ as $Z_{t+1}$ and baseline confounders as $Z_1$.\\
For the causal ordering ($Y_t$, $Z_t$, $X_t$), all the arguments of Supplementary material~\ref{sup_gcomp_yx} hold, so g-computation can be used in the same way as with the reference causal ordering ($Z_t$, $Y_t$, $X_t$). Similarly, all the arguments of Supplementary material~\ref{sup_ccw_yx} hold, so CCW can be used in the same way as with the reference causal ordering ($Z_t$, $Y_t$, $X_t$).
For the causal ordering ($Y_t$, $Z_t$, $X_t$), all the arguments in Supplementary material~\ref{sup_gcomp_yx} apply, so g-computation can be performed as for the reference causal ordering ($Z_t$, $Y_t$, $X_t$). Similarly, all the arguments in Supplementary material~\ref{sup_ccw_yx} apply, so CCW can be performed as for the reference causal ordering ($Z_t$, $Y_t$, $X_t$).\\
For the causal ordering ($X_t$, $Z_t$, $Y_t$), all the arguments in Supplementary material~\ref{sup_gcomp_xy} and Supplementary material~\ref{sup_ccw_xy} apply after relabeling $Z_t$ as $Z_{t+1}$ and baseline confounders as $Z_1$. Thus, g-computation and CCW can be applied as for the reference causal ordering ($Z_t$, $X_t$, $Y_t$) after this relabeling.

\clearpage

\subsection{Supplementary material 5: Identifiability of causal effects in the presence of causal cycles}

In this section we show how causal cycles affect the identifiability of the causal effect and the applicability of CCW.

\subsubsection{Cycle between $X_t$ and $Y_t$}
In the presence of a cycle between the treatment $X$ and the outcome $Y$ as illustrated in Figure~\ref{fig:cycle_yt_xt}, the causal effect is not identifiable. Indeed, Theorem 5 of \cite{ferreiraIdentifyingMacroConditional2025a} shows that the SC-hedge created by the cycle between $X_t$ and $Y_t$ guarantees non-identifiability even if data at a smaller time-step was available.  
\begin{figure}
\centering
\begin{tikzpicture}
\node[state, text={rgb:red,39;green,82;blue,184}, draw={rgb:red,39;green,82;blue,184}, fill=white] at (-9.30, 10.50) (X1) {$X_{1}$};
\node[state, text={rgb:red,39;green,82;blue,184}, draw={rgb:red,39;green,82;blue,184}, fill=white] at (-6.00, 10.50) (X2) {$X_{2}$};
\node[state, text=black, draw=black, fill=white] at (-9.30, 7.50) (Y1) {$Y_{1}$};
\node[state, text={rgb:red,230;green,26;blue,25}, draw={rgb:red,230;green,26;blue,25}, fill=white] at (-6.00, 7.50) (Y2) {$Y_{2}$};
\node[state, text=black, draw=black, fill=white] at (-11.10, 9.00) (Z1) {$Z_{1}$};
\node[state, text=black, draw=black, fill=white] at (-7.80, 9.00) (Z2) {$Z_{2}$};
\path (Y1) edge [bend left=0] (Y2);
\path (X1) edge [bend left=0] (X2);
\path[line width=2.5] (Y1) edge [transform canvas={xshift=4pt}] (X1);
\path[line width=2.5] (X1) edge [transform canvas={xshift=-4pt}] (Y1);
\path (Y1) edge [bend left=-24] (X2);
\path[line width=2.5] (Y2) edge[transform canvas={xshift=6pt}] (X2);
\path[line width=2.5] (X2) edge[transform canvas={xshift=-2pt}] (Y2);
\path (X1) edge [bend left=22] (Y2);
\path (Z1) edge [bend left=0] (X1);
\path (Z1) edge [bend left=0] (Y1);
\path (Z1) edge [bend left=-6] (X2);
\path (Z1) edge [bend left=0] (Y2);
\path (Z2) edge [bend left=0] (X2);
\path (Z2) edge [bend left=0] (Y2);
\path (X1) edge [bend left=0] (Z2);
\path (Z1) edge [bend left=0] (Z2);
\path (Y1) edge [bend left=0] (Z2);
\node[draw=black,dotted,fit=(Z1) (X1) (Y1), inner sep=0.2cm] (box1) {};
\node[above=2pt of box1, align=center] {1st time interval};
\node[draw=black,dotted,fit=(Z2) (X2) (Y2), inner sep=0.2cm] (box2) {};
\node[above=2pt of box2, align=center] {2nd time interval};
\end{tikzpicture}
\caption{Graph with a cycle between $X_t$ and $Y_t$}
\vspace{.5cm}
\label{fig:cycle_yt_xt}
\end{figure}

\subsubsection{Cycle between $X_t$ and $Z_t$}
In the presence of a cycle between the treatment $X$ and the time varying confounders $Z$ as illustrated in Figure~\ref{fig:cycle_zt_xt}, the causal effect is not identifiable. The reason is identical as the one in previous subsection. The cycle between $X_t$ and $Z_t$ still creates a SC-hedge and therefore Theorem 5 of \cite{ferreiraIdentifyingMacroConditional2025a} guarantees non-identifiability even if data at a smaller time-step was available.  
\begin{figure}
\centering
\begin{tikzpicture}
\node[state, text={rgb:red,39;green,82;blue,184}, draw={rgb:red,39;green,82;blue,184}, fill=white] at (-9.30, 10.50) (X1) {$X_{1}$};
\node[state, text={rgb:red,39;green,82;blue,184}, draw={rgb:red,39;green,82;blue,184}, fill=white] at (-6.00, 10.50) (X2) {$X_{2}$};
\node[state, text=black, draw=black, fill=white] at (-9.30, 7.50) (Y1) {$Y_{1}$};
\node[state, text={rgb:red,230;green,26;blue,25}, draw={rgb:red,230;green,26;blue,25}, fill=white] at (-6.00, 7.50) (Y2) {$Y_{2}$};
\node[state, text=black, draw=black, fill=white] at (-11.10, 9.00) (Z1) {$Z_{1}$};
\node[state, text=black, draw=black, fill=white] at (-7.80, 9.00) (Z2) {$Z_{2}$};
\path (Y1) edge [bend left=0] (Y2);
\path (X1) edge [bend left=0] (X2);
\path[line width=2.5] (Z1) edge [transform canvas={xshift=4pt}] (X1);
\path[line width=2.5] (X1) edge [transform canvas={xshift=-4pt}] (Z1);
\path[line width=2.5] (Z2) edge[transform canvas={xshift=6pt}] (X2);
\path[line width=2.5] (X2) edge[transform canvas={xshift=-2pt}] (Z2);
\path (X1) edge [bend left=22] (Y2);
\path (Z1) edge [bend left=0] (Y1);
\path (Z1) edge [bend left=-6] (X2);
\path (Z1) edge [bend left=0] (Y2);
\path (Z2) edge [bend left=0] (Y2);
\path (X1) edge [bend left=0] (Z2);
\path (Z1) edge [bend left=0] (Z2);
\path (Y1) edge [bend left=0] (Z2);
\node[draw=black,dotted,fit=(Z1) (X1) (Y1), inner sep=0.2cm] (box1) {};
\node[above=2pt of box1, align=center] {1st time interval};
\node[draw=black,dotted,fit=(Z2) (X2) (Y2), inner sep=0.2cm] (box2) {};
\node[above=2pt of box2, align=center] {2nd time interval};
\end{tikzpicture}
\caption{Graph with a cycle between $X_t$ and $Z_t$}
\vspace{.5cm}
\label{fig:cycle_zt_xt}
\end{figure}

\subsubsection{Cycle between $Y_t$ and the adjacent time-dependent covariates}

\end{document}